\documentclass[prd,twocolumn,showpacs,amsmath]{revtex4}
\usepackage{graphicx}

\newcommand{\spA}{\quad\ \,}
\newcommand{\spB}{\qquad\qquad}

\newcommand{\bT}{\vec b_T}
\newcommand{\DT}{\vec \Delta_T}
\newcommand{\nT}{\vec 0_T}
\newcommand{\kT}{\vec k_T}
\newcommand{\lT}{\vec l_T}
\newcommand{\pT}{\vec p_T}
\newcommand{\PT}{\vec P_T}
\newcommand{\ppT}{{\vec p\,}'_{\!T}}
\newcommand{\ST}{\vec S_T}

\newcommand{\yT}{\vec y_T}
\newcommand{\zT}{\vec z_T}

\newcommand{\Sop}{\boldsymbol{\hat S}}

\begin{document}
\title{Relations between generalized and\\%
 transverse momentum dependent parton distributions}
\author{S.~Mei{\ss}ner}
\author{A.~Metz}
\author{K.~Goeke}
\affiliation{Institut f{\"u}r Theoretische Physik II,\\%
 Ruhr-Universit{\"a}t Bochum, D-44780 Bochum, Germany}
\date{\today}
\begin{abstract}
Recent work suggests nontrivial relations between generalized parton distributions 
on the one hand and (naive time-reversal odd) transverse momentum dependent 
distributions on the other.
Here we review the present knowledge on such type of relations.
Moreover, as far as spectator model calculations are concerned, the existing 
results are considerably extended.
While various relations between the two types of parton distributions can be found
in the framework of spectator models, so far no nontrivial model-independent 
relations have been established.
\end{abstract}
\maketitle

\section{Introduction}
For decades the partonic structure of the nucleon was almost entirely discussed 
in terms of unpolarized and polarized forward parton densities, which merely depend 
on the longitudinal momentum of the respective parton.
A much more comprehensive picture of the nucleon structure, however, can be obtained 
by considering two other, more general types of parton distributions:
first, generalized parton distributions (GPDs) entering the QCD description of hard 
exclusive reactions on the nucleon 
(see, e.g., Refs.~\cite{Mueller:1998fv,Diehl:2003ny,Belitsky:2005qn});
second, transverse momentum dependent parton distributions (TMDs) entering the 
description of various hard semi-inclusive reactions 
(see, e.g., Ref.~\cite{Barone:2001sp,Mulders:1995dh,Bacchetta:2006tn}).
Not only a large body of theoretical work on these types of parton correlators 
appeared during the last decade, but also new high luminosity particle accelerators 
nowadays allow one to explore such intriguing though complicated objects 
experimentally.

On the TMD side the so-called naive time-reversal odd (T-odd) parton distributions
are of particular importance, because these objects can give rise to single spin 
asymmetries (SSAs).
Single spin phenomena were measured in hadron-hadron collisions at 
FermiLab~\cite{Adams:1991rw,Adams:1991cs} and at 
RHIC~\cite{Adams:2003fx,Adler:2005in}, as well as in lepton-hadron collisions by the 
COMPASS Collaboration~\cite{Alexakhin:2005iw,Ageev:2006da}, the HERMES 
Collaboration~\cite{Airapetian:1999tv,Airapetian:2001eg,Airapetian:2002mf,Airapetian:2004tw,Airapetian:2005jc,Diefenthaler:2005gx,Airapetian:2006rx}, 
and at JeffersonLab~\cite{Avakian:2003pk}.
In general, the theoretical description of such observables in the framework of QCD 
has been and still is a challenge for the theory.
However, a crucial step forward was the observation that time-reversal invariance of 
the strong interaction does not forbid the existence of 
T-odd TMDs~\cite{Collins:2002kn,Brodsky:2002cx}. 
At least for the SSAs in lepton-induced reactions a QCD-factorization formula containing 
T-odd TMDs has been put forward in the 
meantime~\cite{Ji:2004wu,Ji:2004xq,Collins:2004nx}.

An important object in this context is the T-odd 
Sivers function~\cite{Sivers:1989cc,Sivers:1990fh}, denoted by $f_{1T}^\perp$ in the
nomenclature of Ref.~\cite{Boer:1997nt}.
The Sivers function quantifies the SSA related to the transverse momentum 
dependent distribution of unpolarized partons inside a transversely polarized target.
Experimentally $f_{1T}^\perp$ can be studied, e.g., by measuring a transverse 
SSA in semi-inclusive deep-inelastic scattering (SIDIS).
Extractions of the Sivers function from existing SIDIS 
data~\cite{Airapetian:2004tw,Diefenthaler:2005gx,Alexakhin:2005iw,Ageev:2006da} 
have already been 
performed~\cite{Efremov:2004tp,Anselmino:2005nn,Anselmino:2005ea,Vogelsang:2005cs,Collins:2005ie,Anselmino:2005an}.

An intuitive picture of various transverse SSAs, in particular, also of the 
Sivers effect, was proposed in Refs.~\cite{Burkardt:2002ks,Burkardt:2003uw}.
That work suggested for the first time that there may be a close connection between 
a certain GPD, typically denoted by $E$ (see, e.g., Ref.~\cite{Diehl:2003ny}), 
and the Sivers asymmetry.
In order to generate the Sivers effect in SIDIS two ingredients are required in the 
picture of Refs.~\cite{Burkardt:2002ks,Burkardt:2003uw}:
first, the impact parameter distribution of unpolarized quarks in a transversely 
polarized target has to be distorted.
Such a distortion is directly connected to the GPD $E$.
Second, the fragmenting struck quark in SIDIS has to experience a final state 
interaction with the target spectators. 

Though the work in~\cite{Burkardt:2002ks,Burkardt:2003uw} provided a very attractive 
explanation of the origin of the Sivers effect, it did not provide a quantitative
connection between the GPD $E$ and $f_{1T}^\perp$.
For a particular moment of the Sivers function such a connection was later on 
established in a perturbative low order calculation using a simple diquark spectator 
model of the nucleon~\cite{Burkardt:2003je}.
Another step forward was made by comparing the correlator for chiral-odd GPDs 
in impact parameter space on the one hand with the correlator for chiral-odd TMDs 
on the other~\cite{Diehl:2005jf}.
The work~\cite{Diehl:2005jf} suggested, in particular, a relation between the second 
leading twist T-odd quark TMD, the Boer-Mulders function $h_1^\perp$~\cite{Boer:1997nt}, 
and a certain linear combination of GPDs.
This possible connection between $h_1^\perp$ and GPDs was afterwards discussed in 
more detail in Ref.~\cite{Burkardt:2005hp}.
In the framework of the diquark spectator model quite recently another relation
between the GPD $E$ and the Sivers function was obtained~\cite{Lu:2006kt}, which is 
similar to the one found in~\cite{Burkardt:2003je}.

Despite these developments so far no nontrivial model-independent relations between 
GPDs and TMDs have been established.
(An attempt in this direction was made, e.g., in Ref.~\cite{Burkardt:2003uw}.)
If one considers for instance the Sivers function, the crucial problem lies in the
fact that there is apparently no model-independent factorization of $f_{1T}^\perp$ 
into a distortion effect, described by the GPD $E$ in impact parameter space, 
times a final state interaction.

The manuscript is organized in the following way.
In Sec.~\ref{s:two} we give our definitions of the GPDs, both in momentum and 
in impact parameter space, and of the TMDs.
In particular, we also include the gluon sector which so far in the literature
has not been discussed at all in the context of possible relations between GPDs 
and TMDs.
Section~\ref{s:three} is devoted to model-independent considerations of the
relations between GPDs and TMDs.
Here the current knowledge on this point is summarized and the main difficulties
are presented.
Also new possible relations for gluon parton distributions are provided, which  
later on in the manuscript are investigated in a model calculation.
Section~\ref{s:four} describes model results on relations between GPDs and TMDs, 
where we exploit two perturbative models for the target: 
the scalar diquark spectator model of the nucleon,
and a quark target model treated in perturbative QCD.
The latter, in particular, allows one to study the gluon sector.
The model results of the various parton distributions, calculated to lowest 
nontrivial order in perturbation theory, show relations in accordance with
the considerations in Sec.~\ref{s:three}. 
In the framework of the two models also on the quark sector new relations 
are established.
Some of these relations contain the results of Ref.~\cite{Burkardt:2003je} 
and of Ref.~\cite{Lu:2006kt} as limiting cases.
In this section we also argue that even in the context of spectator models  
certain nontrivial relations between GPDs and TMDs are far from being obvious 
if one considers higher orders in perturbation theory.
Our summary is given in Sec.~\ref{s:five}.
The results on the various parton distributions in the two target models are 
collected in two appendices.

\section{Definition of parton distributions}
\label{s:two}
\subsection{Generalized parton distributions (GPDs)}
\begin{figure}[t]
 \includegraphics{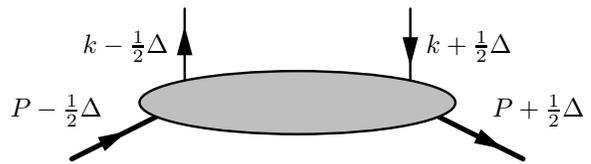}
 \caption{Kinematics for GPDs.}
 \label{f:1}
\end{figure}

We start by presenting the definitions of the GPDs.
Unless stated otherwise we follow here the conventions of 
Ref.~\cite{Diehl:2003ny}.
The momenta of the incoming and outgoing nucleon are given by 
(see also Fig.~\ref{f:1})
\begin{equation}
 p=P-\tfrac{1}{2}\Delta \,, \qquad
 p'=P+\tfrac{1}{2}\Delta \,,
\end{equation}
and satisfy $p^2 = p'^2 = M^2$, with $M$ denoting the nucleon mass.
The GPDs depend on the three variables
\begin{equation}
 x=\frac{k^+}{P^+} \,, \qquad
 \xi=-\frac{\Delta^+}{2P^+} \,, \qquad
 t=\Delta^2 \,,
\end{equation}
where the light-cone coordinates are defined by
\begin{equation}
 v^\pm=\tfrac{1}{\sqrt{2}}(v^0\pm v^3) \,, \qquad
 \vec{v}_T=(v^1,v^2)
\end{equation}
for a generic 4-vector $v$.
In a physical process the so-called skewness $\xi$ and the momentum transfer $t$ 
to the nucleon are fixed by the external kinematics, whereas $x$ is typically an 
integration variable.
It is convenient to define the following tensors,
\begin{equation}
 \delta_T^{ij}=-g^{ij} \,, \qquad
 \epsilon_T^{ij}=\epsilon^{-+ij} \,.
\end{equation}
Moreover, we use the conventions
\begin{equation}
 \epsilon^{0123}=1 \,,
\end{equation}
\begin{equation}
 \gamma_5=i\gamma^0\gamma^1\gamma^2\gamma^3 \,,
\end{equation}
\begin{equation}
 \sigma^{\mu\nu}=\tfrac{i}{2}(\gamma^\mu\gamma^\nu-\gamma^\nu\gamma^\mu) \,.
\end{equation}

The quark GPDs are defined through the light-cone correlation function
\begin{align} \label{e:qgpd}
 &F^{q[\Gamma]}(x,\Delta;\lambda,\lambda')\notag\\
 &\ =\frac{1}{2}\int\frac{dz^-}{2\pi}\,e^{i k\cdot z}\,
  \big<p';\lambda'\big|\,
  \bar\psi\big(\!-\!\tfrac{1}{2}z\big)\,\Gamma\notag\\
 &\spA\times\mathcal{W}\big(\!-\!\tfrac{1}{2}z;\tfrac{1}{2}z\big)\,
  \psi\big(\tfrac{1}{2}z\big)\,\big|p;\lambda\big>\,
  \Big|_{\substack{z^+=0^+\\\zT=\nT}} \,,
\end{align}
where $\lambda$ and $\lambda'$ respectively characterize the helicity of the 
nucleon in the initial and final state.
The object $\Gamma$ is a generic matrix in Dirac space.
In~(\ref{e:qgpd}) a summation over the color of the quark fields is understood. 
Furthermore, in a perturbative calculation of the correlation function only 
connected graphs have to be taken into account.
The correlation function in~(\ref{e:qgpd}) also depends on a renormalization
scale which we disregard throughout this work.
The Wilson line $\mathcal{W}$ in~(\ref{e:qgpd}), connecting the two quark fields
and ensuring color gauge invariance of the correlator, is given by
\begin{align}
 &\mathcal{W}\big(\!-\!\tfrac{1}{2}z;\tfrac{1}{2}z\big)
             \Big|_{\substack{z^+=0^+\\\zT=\nT}}\notag\\
 &\ =\big[0^+,-\tfrac{1}{2}z^-,\nT;0^+,\tfrac{1}{2}z^-,\nT\big]\notag\\
 &\ =\mathcal{P}\exp\bigg(\!-i g\int_{-(1/2)z^-}^{(1/2)z^-}\!\!\!dy^-\,
  t_a\,A^+_a\big(0^+,y^-,\nT\big)\bigg) \,,
\end{align}
with $\mathcal{P}$ denoting path-ordering and $t_a$ representing the Gell-Mann 
matrices.
For three particular matrices $\Gamma$ in (\ref{e:qgpd}) one obtains the leading 
twist (twist-2) GPDs:
\begin{align} \label{e:qgpd1}
 &F^{q}(x,\Delta;\lambda,\lambda')
  =F^{q[\gamma^+]}(x,\Delta;\lambda,\lambda')\notag\\
 &\ =\frac{1}{2P^+}\,\bar u(p',\lambda')\,
  \bigg(\gamma^+\,H^{q}(x,\xi,t)\notag\\
 &\spB+\frac{i\sigma^{+\mu}\Delta_\mu}{2M}\,E^{q}(x,\xi,t)\bigg)\,
  u(p,\lambda) \,, 
  \displaybreak[0]\\ \label{e:qgpd2}
 &\tilde F^{q}(x,\Delta;\lambda,\lambda')
  =F^{q[\gamma^+\gamma_5]}(x,\Delta;\lambda,\lambda')\notag\\
 &\ =\frac{1}{2P^+}\,\bar u(p',\lambda')\,
  \bigg(\gamma^+\gamma_5\,\tilde H^{q}(x,\xi,t)\notag\\
 &\spB+\frac{\Delta^+\gamma_5}{2M}\,\tilde E^{q}(x,\xi,t)\bigg)\,
  u(p,\lambda) \,, 
 \displaybreak[0]\\ \label{e:qgpd3}
 &F_T^{q,\,j}(x,\Delta;\lambda,\lambda')
  =F^{q[i\sigma^{j+}\gamma_5]}(x,\Delta;\lambda,\lambda')\notag\\
 &\ =-\frac{i\epsilon_T^{ij}}{2P^+}\,\bar u(p',\lambda')\,
  \bigg(i\sigma^{+i}\,H_T^{q}(x,\xi,t)\notag\\
 &\spB+\frac{\gamma^+\Delta_T^i-\Delta^+\gamma_T^i}{2M}\,E_T^{q}(x,\xi,t)\notag\\
 &\spB+\frac{P^+\Delta_T^i-\Delta^+P_T^i}{M^2}\,\tilde H_T^{q}(x,\xi,t)\notag\\
 &\spB+\frac{\gamma^+P_T^i-P^+\gamma_T^i}{M}\,\tilde E_T^{q}(x,\xi,t)
  \bigg)\,u(p,\lambda) \,.
\end{align}
In the so-called chiral-odd sector in Eq.~(\ref{e:qgpd3}) one may equally well 
work with $\Gamma = \sigma^{j+} = - \epsilon_T^{jk} \, i \sigma^{k+} \gamma_5$. 

Notice that the GPDs for antiquarks are defined analogously to the quark GPDs.
One merely has to replace in~(\ref{e:qgpd}) the quark fields by the corresponding 
charge-conjugated fields.
Unless stated otherwise all results in the following apply also to the case of 
distributions for antiquarks.

As mentioned in the introduction we also want to consider possible relations
for gluon distributions.
The relevant correlation function for leading twist gluon GPDs 
reads~\cite{Diehl:2003ny}
\begin{align} \label{e:ggpd}
 &F^{g[ij]}(x,\Delta;\lambda,\lambda')\notag\\
 &\ =\frac{1}{xP^+}\int\frac{dz^-}{2\pi}\,e^{i k\cdot z}\,
  \big<p';\lambda'\big|\,
  F^{+j}_a\big(\!-\!\tfrac{1}{2}z\big)\notag\\
 &\spA\times\mathcal{W}_{ab}\big(\!-\!\tfrac{1}{2}z;\tfrac{1}{2}z\big)\,
  F^{+i}_b\big(\tfrac{1}{2}z\big)\,\big|p;\lambda\big>\,
  \Big|\,\!_{\substack{z^+=0^+\\\zT=\nT}} \,,
\end{align}
where the Wilson line is given in the adjoint representation of the color 
$SU(3)$,
\begin{align}
 &\!\mathcal{W}_{ab}\big(\!-\!\tfrac{1}{2}z;\tfrac{1}{2}z\big)
  \Big|_{\substack{z^+=0^+\\\zT=\nT}}\notag\\
 &\!\ =\big[0^+,-\tfrac{1}{2}z^-,\nT;0^+,\tfrac{1}{2}z^-,\nT\big]_{ab}\notag\\
 &\!\ =\mathcal{P}\exp\bigg(\!-g\int_{-(1/2)z^-}^{(1/2)z^-}\!\!\!dy^-\,
  f_{abc}\,A^+_c\big(0^+,y^-,\nT\big)\bigg) \,.\!
\end{align}
The gluon field strength tensor in~(\ref{e:ggpd}) has the standard form
\begin{equation} \label{e:fqcd}
 F^{\mu\nu}_a(x)=\partial^\mu A^\nu_a(x)-\partial^\nu A^\mu_a(x)
                + gf_{abc}\,A^\mu_b(x)\,A^\nu_c(x) \,,
\end{equation}
with $f_{abc}$ being the structure constants of the $SU(3)$.
For the definition of the chiral-odd gluon GPDs we will need the symmetry 
operator $\Sop$ defined through
\begin{equation}
 \Sop\,O^{ij}=\tfrac{1}{2}\big(O^{ij}+O^{ji}-\delta_T^{ij}\,O^{mm}\big) 
\end{equation}
for a generic tensor $O^{ij}$.
One readily observes that the symmetrized tensor $\Sop\,O^{ij}$ has only two
independent components.
The twist-2 gluon GPDs are given through the correlator in Eq.~(\ref{e:ggpd}) 
according to
\begin{align} \label{e:ggpd1}
 &F^g(x,\Delta;\lambda,\lambda')
  =\delta_T^{ij}\,F^{g[ij]}(x,\Delta;\lambda,\lambda')\notag\\
 &\ =\frac{1}{2P^+}\,\bar u(p',\lambda')\,
  \bigg(\gamma^+\,H^g(x,\xi,t)\notag\\
 &\spB+\frac{i\sigma^{+\mu}\Delta_\mu}{2M}\,E^g(x,\xi,t)\bigg)\,
  u(p,\lambda) \,, 
 \displaybreak[0]\\ \label{e:ggpd2}
 &\tilde F^g(x,\Delta;\lambda,\lambda')
  =i\epsilon_T^{ij}\,F^{g[ij]}(x,\Delta;\lambda,\lambda')\notag\\
 &\ =\frac{1}{2P^+}\,\bar u(p',\lambda')\,
  \bigg(\gamma^+\gamma_5\,\tilde H^g(x,\xi,t)\notag\\
 &\spB+\frac{\Delta^+\gamma_5}{2M}\,\tilde E^g(x,\xi,t)\bigg)\,
  u(p,\lambda) \,, 
 \displaybreak[0]\\ \label{e:ggpd3}
 &F_T^{g,\,ij}(x,\Delta;\lambda,\lambda')
  =-\Sop\,F^{g[ij]}(x,\Delta;\lambda,\lambda')\notag\\
 &\ =\frac{\Sop}{2P^+}\,\frac{P^+\Delta_T^i-\Delta^+P_T^i}{2MP^+}\,
  \bar u(p',\lambda')\,\bigg(i\sigma^{+j}\,H_T^g(x,\xi,t)\notag\\
 &\spB+\frac{\gamma^+\Delta_T^j-\Delta^+\gamma_T^j}{2M}\,E_T^g(x,\xi,t)\notag\\
 &\spB+\frac{P^+\Delta_T^j-\Delta^+P_T^j}{M^2}\,\tilde H_T^g(x,\xi,t)\notag\\
 &\spB+\frac{\gamma^+P_T^j-P^+\gamma_T^j}{M}\,\tilde E_T^g(x,\xi,t)
  \bigg)\,u(p,\lambda) \,.
\end{align}
Note that the definitions of the chiral-even quark and gluon GPDs directly 
correspond to each other [compare the right-hand side (RHS) of~(\ref{e:qgpd1}) and~(\ref{e:ggpd1}), 
as well as the RHS of~(\ref{e:qgpd2}) and~(\ref{e:ggpd2})].
On the other hand, in the chiral-odd sector the definitions of the quark and gluon 
GPDs are (symbolically) connected by 
\begin{equation}
 F_T^{g,\,ij} \leftrightarrow 
 i\Sop\,\frac{P^+\Delta_T^i-\Delta^+P_T^i}{2MP^+}\,\epsilon_T^{jk}F_T^{q,\,k} 
\end{equation}
[compare the RHS of~(\ref{e:qgpd3}) and~(\ref{e:ggpd3})].
We also mention that our definition of all gluon GPDs $X^g$ differs by a factor $x$ 
from the one of Ref.~\cite{Diehl:2003ny},
\begin{equation} \label{e:gpdgluondef}
X^g(x,\xi,t)\Big|_{\textrm{here}} = 
\frac{1}{x} X^g(x,\xi,t)\Big|_{\textrm{Ref.~\cite{Diehl:2003ny}}} \,.
\end{equation}
The main advantage of this choice in the context of our work is that the structure of 
the relations between GPDs and TMDs for quarks and gluons will look alike.

Altogether there exist eight leading twist quark GPDs and eight leading twist 
gluon GPDs.
All GPDs are real-valued which follows from time-reversal.
Moreover, using commutation relations for the parton fields, they obey symmetry 
relations of the type 
\begin{equation} 
 X^{\bar q/g}(x,\xi,t)=\pm\,X^{q/g}(-x,\xi,t) \,,
\end{equation}
where the \emph{minus} holds for all GPDs $X$ except $\tilde H$ and $\tilde E$. 
It is therefore sufficient to consider only the region $x>0$ as we will do in 
the present work.
Eventually, hermiticity implies
\begin{equation} \label{e:xirel}
 X^{q/g}(x,\xi,t)=\pm\,X^{q/g}(x,-\xi,t) \,,
\end{equation}
where the \emph{plus} holds for all GPDs $X$ except $\tilde E_T$.
The relation~(\ref{e:xirel}) is needed later on in order to write down
the general structure of the GPD correlator for $\xi = 0$.

\subsection{GPDs in impact parameter space}
In a next step we want to consider the GPDs in transverse position
(impact parameter) 
space~\cite{Burkardt:2000za,Ralston:2001xs,Diehl:2002he,Burkardt:2002hr}.
Of particular interest is the case $\xi = 0$, where a density interpretation 
of GPDs in impact parameter space may be obtained~\cite{Burkardt:2000za}.
Such an interpretation, in principle, allows one to study a three-dimensional picture 
of the nucleon.
In the context of the present work the impact parameter picture is relevant 
for different reasons:
first, the intuitive picture for various transverse SSAs in semi-inclusive 
processes given in~\cite{Burkardt:2002ks,Burkardt:2003uw} is based on the impact 
parameter representation of the GPD $E^q$.
Second, the quantitative relation between the Sivers function $f_{1T}^{\bot q}$ 
and the GPD $E^q$, obtained in the framework of a scalar diquark spectator model of 
the nucleon~\cite{Burkardt:2003je}, also contains $E^q$ in impact parameter space.
Third, the impact parameter representation was used to point out analogies 
between chiral-odd quark GPDs and TMDs~\cite{Diehl:2005jf}.
Fourth, we use this representation as a guidance to obtain new possible 
relations between GPDs and TMDs in the gluon sector.

For the following discussion it is convenient to introduce a state describing
the incoming nucleon with both longitudinal and transverse polarization as a 
superposition of states with definite light-cone helicity~\cite{Diehl:2005jf},
\begin{equation} \label{e:spingpd}
\big|p;S\big> = \cos(\tfrac{1}{2}\vartheta) \, \big|p;+\big> +
             \sin(\tfrac{1}{2}\vartheta) \, e^{i\varphi} \, \big|p;-\big> \,.
\end{equation}
If one transforms this state to the rest frame of the nucleon, it describes a particle 
whose three-dimensional spin vector is given by
\begin{align}
\vec{S} &= (S^1,\, S^2,\,S^3)
 \notag \\ 
 &= (\sin\vartheta \, \cos\varphi,\, \sin\vartheta \, \sin\varphi,\, \cos\vartheta) \,,
\end{align}
see Ref.~\cite{Diehl:2005jf}.
In the following we use the notation $\lambda = S^3 = \cos\vartheta$.
By means of the definition in~(\ref{e:spingpd}) a corresponding state $\big<p';S\big|$ 
for the outgoing nucleon (with the same spin vector $\vec{S}$) can be specified.
Replacing now the helicity states in the correlators for the quark and gluon 
GPDs according to
\begin{equation}
\big<p';\lambda'\big| \to \big<p';S\big| \,, \qquad
\big|p;\lambda\big> \to \big|p;S\big> \,, 
\end{equation}
the matrix elements for all possible helicity combinations can be obtained 
by an appropriate choice of the spin vector.
This statement is obvious looking at the relation 
\begin{align}
 &F(x,\Delta;S)\notag\\
 &\ =\tfrac{1}{2}\,\big[F(x,\Delta,+,+)+F(x,\Delta,-,-)\big]\notag\\
 &\spA+\tfrac{1}{2}\lambda\,\big[F(x,\Delta,+,+)-F(x,\Delta,-,-)\big]\notag\\
 &\spA+\tfrac{1}{2}S_T^1\,\big[F(x,\Delta,-,+)+F(x,\Delta,+,-)\big]\notag\\
 &\spA+\tfrac{i}{2}S_T^2\,\big[F(x,\Delta,-,+)-F(x,\Delta,+,-)\big]\,,
\end{align}
which can be readily verified.
Before considering the transformation to the impact parameter space we also give
the definition of the light-cone helicity spinors.
The calculations are performed using the conventions of Ref.~\cite{Lepage:1980fj}, 
\begin{align}
 u(p,+)&=\frac{1}{\sqrt{2^{3/2}\,p^+}}\,
 \begin{pmatrix}
  \sqrt{2}\,p^++m_q\\
  \quad\ p_T^1+i p_T^2\quad\ \\
  \sqrt{2}\,p^+-m_q\\
  p_T^1+i p_T^2
 \end{pmatrix} \,, \displaybreak[0]\\
 u(p,-)&=\frac{1}{\sqrt{2^{3/2}\,p^+}}\,
 \begin{pmatrix}
  -p_T^1+i p_T^2\\
  \sqrt{2}\,p^++m_q\\
  \quad\ p_T^1-i p_T^2\quad\ \\
  -\sqrt{2}\,p^++m_q 
 \end{pmatrix} \,.
\end{align}

When defining the GPDs in impact parameter space we restrict ourselves to the case
$\xi = 0$ for which a density interpretation exists~\cite{Burkardt:2000za}. 
For convenience we also use $\PT=\nT$.
These two conditions imply $p^+ = p'^+ = P^+$ and $\Delta^+ = \Delta^- = 0$.
The parton correlators in impact parameter space are now given by the Fourier transform
\begin{equation} \label{e:corrimpact}
 \mathcal{F}(x,\bT;S)=\int\frac{d^2\DT}{(2\pi)^2}\,e^{-i\DT\cdot\bT}\,F(x,\Delta_T;S) \,.
\end{equation}
The impact parameter $\vec{b}_T$ and the transverse part of the momentum transfer 
$\vec{\Delta}_T$ are conjugate variables.
In the impact parameter representation one naturally obtains diagonal matrix elements
which is the crucial prerequisite for a density interpretation.
This gain of the impact parameter picture becomes evident after introducing
the states~\cite{Soper:1976jc,Burkardt:2000za,Diehl:2002he}
\begin{align} \label{e:inistate}
 \big|P^+,\bT;S\big>&=\mathcal{N}\int\frac{d^2\pT}{(2\pi)^2}\,e^{-i\pT\cdot\bT}\,\big|p;S\big> \,, 
 \\ \label{e:outstate}
 \big<P^+,\bT;S\big|&=\mathcal{N^*}\int\frac{d^2\ppT}{(2\pi)^2}\,e^{i\ppT\cdot\bT}\,\big<p';S\big| \,,
\end{align}
which characterize a nucleon with momentum $P^+$ at a transverse position $\vec{b}_T$ 
and a polarization specified by $S$.
The normalization factor $\mathcal{N}$ in these formulas is given by 
\begin{equation}
 \frac{1}{|\mathcal{N}|^2}=\int\frac{d^2\pT}{(2\pi)^2}
\end{equation}
and therefore infinite.
However, using wave packets instead of plane wave states this infinity can be 
avoided~\cite{Burkardt:2000za,Diehl:2002he}.
With the states in Eqs.~(\ref{e:inistate}) and~(\ref{e:outstate}) the correlators defining the 
GPDs of quarks and gluons can be rewritten as
\begin{align} \label{e:qimpact}
 &\mathcal{F}^{q[\Gamma]}(x,\bT;S)\notag\\
 &\ =\frac{1}{2}\int\frac{dz^-}{2\pi}\,e^{ixP^+z^-}\,
  \big<P^+,\nT;S\big|\,
  \bar\psi\big(z_1\big)\,\Gamma\notag\\
 &\spA\times\mathcal{W}\big(z_1;z_2\big)\,
  \psi\big(z_2\big)\,\big|P^+,\nT;S\big> \,, \\
 &\mathcal{F}^{g[ij]}(x,\bT;S)\notag\\ \label{e:gimpact}  
 &\ =\frac{1}{xP^+}\int\frac{dz^-}{2\pi}\,e^{ixP^+z^-}\,
  \big<P^+,\nT;S\big|\,
  F^{+j}_a\big(z_1\big)\notag\\
 &\spA\times\mathcal{W}_{ab}\big(z_1;z_2\big)\,
  F^{+i}_b\big(z_2\big)\,\big|P^+,\nT;S\big> \,,
\end{align}
with
\begin{equation} \label{e:zdef}
 z_{1/2}=(0^+,\mp\tfrac{1}{2}z^-,\bT) \,.
\end{equation}
Obviously, the two correlation functions in~(\ref{e:qimpact}) and~(\ref{e:gimpact}) are 
diagonal. 

In analogy with Eq.~(\ref{e:corrimpact}) we define the GPDs in impact parameter space 
according to
\begin{equation}
 \mathcal{X}(x,\bT^{\,2})=\int\frac{d^2\DT}{(2\pi)^2}\,e^{-i\DT\cdot\bT}\,X(x,0,-\DT^2) \,.
\end{equation}
Using this definition one finds after straightforward algebra that the correlators 
in Eqs.~(\ref{e:qgpd1})--(\ref{e:qgpd3}) for the quarks and in 
Eqs.~(\ref{e:ggpd1})--(\ref{e:ggpd3}) for the gluons, written in impact parameter space 
at the kinematical point $\xi = 0$, take the form 
\begin{align} \label{e:impact1}
 &\!\mathcal{F}^{q/g}(x,\bT;S)\notag\\
 &\!\ =\mathcal{H}^{q/g}(x,\bT^{\,2})
  +\frac{\epsilon_T^{ij}b_T^iS_T^j}{M}\,\bigg(\mathcal{E}^{q/g}(x,\bT^{\,2})\bigg)' \,,
  \displaybreak[0]\\ \label{e:impact2}
 &\!\mathcal{\tilde F}^{q/g}(x,\bT;S)\notag\\
 &\!\ =\lambda\,\mathcal{\tilde H}^{q/g}(x,\bT^{\,2}) \,,
  \displaybreak[0]\\ \label{e:impact3}
 &\!\mathcal{F}_T^{q,\,j}(x,\bT;S) \notag\\
 &\!\ =\frac{\epsilon_T^{ij}b_T^i}{M}\,\bigg(\mathcal{E}_T^{q}(x,\bT^{\,2})
  +2\mathcal{\tilde H}_T^{q}(x,\bT^{\,2})\bigg)'\notag\\
 &\!\spA+S_T^j\,\bigg(\mathcal{H}_T^{q}(x,\bT^{\,2})
  -\frac{\bT^{\,2}}{M^2}\,\Delta_b\mathcal{\tilde H}_T^{q}(x,\bT^{\,2})\bigg)\notag\\
 &\!\spA+\frac{2b_T^j\,\bT\cdot\ST-S_T^j\,\bT^{\,2}}{M^2}\,
  \bigg(\mathcal{\tilde H}_T^{q}(x,\bT^{\,2})\bigg)'' \,,
  \displaybreak[0]\\ \label{e:impact4}
 &\!\mathcal{F}_T^{g,\,ij}(x,\bT;S) \notag\\
 &\!\ =-\frac{\Sop\,b_T^ib_T^j}{M^2}\,
  \bigg(\mathcal{E}_T^g(x,\bT^{\,2})
  +2\mathcal{\tilde H}_T^g(x,\bT^{\,2})\bigg)''\notag\\
 &\!\spA-\frac{\Sop\,b_T^i\epsilon_T^{jk}S_T^k}{M}\,
  \bigg(\mathcal{H}_T^{g}(x,\bT^{\,2})-
  \frac{\bT^{\,2}}{M^2}\,\Delta_b\mathcal{\tilde H}_T^g(x,\bT^{\,2})\bigg)'\notag\\
 &\!\spA-\frac{\Sop\,b_T^i\epsilon_T^{jk}
  \big(2b_T^k\,\bT\cdot\ST-S_T^k\,\bT^{\,2}\big)}{M^3}\,
  \bigg(\mathcal{\tilde H}_T^g(x,\bT^{\,2})\bigg)'''\!\!\! \,. \!\!\!
\end{align}
In these equations we use the notation 
\begin{equation}
  \Big(\mathcal{X}(x,\bT^{\,2})\Big)'
  =\frac{\partial}{\partial\bT^{\,2}} \, \bigg(\mathcal{X}(x,\bT^{\,2})\bigg) \,,
\end{equation}
and analogous for the higher derivatives of the GPDs $\mathcal{X}$, as well as
\begin{equation} \label{e:gpdabbr}
  \Delta_b\mathcal{X}(x,\bT^{\,2})
 =\frac{1}{\bT^{\,2}}\,\frac{\partial}{\partial\bT^{\,2}}
  \bigg[\bT^{\,2}\,\frac{\partial}{\partial\bT^{\,2}} \, 
  \bigg(\mathcal{X}(x,\bT^{\,2})\bigg)\bigg] \,.
\end{equation}
While Eqs.~(\ref{e:impact1})--(\ref{e:impact3}) were already given in the 
literature~\cite{Burkardt:2002ks,Diehl:2005jf}, the result in~(\ref{e:impact4}) is new.
Since the point $\xi = 0$ is chosen, the GPDs $\tilde{E}$ and $\tilde{E}_T$ 
do not show up in (\ref{e:impact2})--(\ref{e:impact4}):
the GPD $\tilde{E}$ is multiplied by the kinematical factor $\Delta^+ = 0$ 
in the correlator, and $\tilde{E}_T$ vanishes due to the constraint in Eq.~(\ref{e:xirel}).

The expression in (\ref{e:impact1}), for instance, can be interpreted as the density of 
unpolarized quarks/gluons with momentum fraction $x$ at the transverse position $\vec{b}_T$ 
in a (transversely polarized) proton.
This density has a spin-independent part given by $\mathcal{H}$, and a spin-dependent 
part proportional to the derivative of $\mathcal{E}$.
Some details on the physical interpretation of~(\ref{e:impact2}) and~(\ref{e:impact3}) can be 
found in Refs.~\cite{Diehl:2005jf,Burkardt:2005hp}.

Because of the spin-dependent term the impact parameter distribution in~(\ref{e:impact1}) 
is not axially symmetric (unless $\mathcal{E}'  = 0$), i.e., it depends on the direction 
of $\vec{b}_T$.
In other words, the spin part causes a distortion of the distribution~(\ref{e:impact1}).
Note that the RHS in~(\ref{e:impact3}) contains two terms providing a distortion, 
one determined by the first derivative of $\mathcal{E}_T + 2\mathcal{\tilde{H}}_T$ and one 
given by the second derivative of $\mathcal{\tilde{H}}_T$.
In~(\ref{e:impact4}) none of the three terms on the RHS is axially symmetric.
Later on, we will use the results~(\ref{e:impact1})--(\ref{e:impact4}) and compare them with 
the corresponding correlators for TMDs. 
This procedure will give us some guidance in order to obtain possible relations between 
GPDs and TMDs (see also Ref.~\cite{Diehl:2005jf}).
It turns out that the specific form of the relations depends on the number of derivatives 
of the GPDs in Eqs.~(\ref{e:impact1})--(\ref{e:impact4}).

\subsection{Transverse momentum dependent\\parton distributions (TMDs)}
\begin{figure}[t]
 \includegraphics{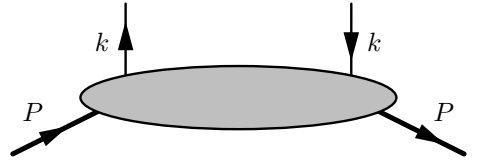}
 \caption{Kinematics for TMDs.}
 \label{f:2}
\end{figure}
In this subsection we summarize our notation for the TMDs.
For the quark sector we follow the conventions of 
Refs.~\cite{Mulders:1995dh,Boer:1997nt,Goeke:2005hb,Bacchetta:2006tn}, while for the 
gluons our treatment is similar to Ref.~\cite{Mulders:2000sh}, where gluon TMDs were  
systematically classified for the first time.
In Fig.~\ref{f:2} the kinematics for TMDs is indicated.
The nucleon momentum is denoted by $P$ (with $P^2 = M^2$), the quark momentum 
by $k$. 
The TMDs depend on $k^+$, given by the longitudinal momentum fraction
\begin{equation}
 x=\frac{k^+}{P^+} \,, \qquad
\end{equation}
and on the transverse momentum $\vec{k}_T$.
Like in the case of GPDs a possible polarization of the nucleon can conveniently be
described by a covariant spin vector $S$, whose components read
\begin{equation}
 S^+=\frac{\lambda P^+}{M} \,, \qquad
 S^-=-\frac{\lambda P^-}{M} \,,\qquad
 \ST \,.
\end{equation}

\begin{figure}
 \includegraphics{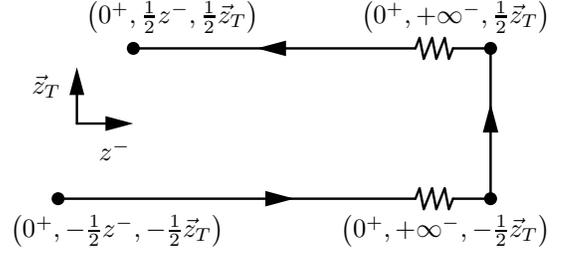}
 \caption{Path of Wilson line for TMDs in SIDIS.}
 \label{f:3}
\end{figure}

The quark TMDs are defined through the correlation function
\begin{align} \label{e:qtmd}
 &\Phi^{q[\Gamma]}(x,\kT;S)\notag\\
 &\ =\frac{1}{2}\int\frac{dz^-}{2\pi}\,\frac{d^2\zT}{(2\pi)^2}\,e^{i k\cdot z}\,
  \big<P;S\big|\,
  \bar\psi\big(\!-\!\tfrac{1}{2}z\big)\,\Gamma\notag\\
 &\spA\times\mathcal{W}_{+\infty}\big(\!-\!\tfrac{1}{2}z;\tfrac{1}{2}z\big)\,
  \psi\big(\tfrac{1}{2}z\big)\,\big|P;S\big>\,
  \Big|_{z^+=0^+} \,,
\end{align}
in which a summation over the color of the quark fields is implicit, and only connected 
diagrams have to be considered in perturbative calculations.
Like in the case of GPDs, the dependence of the correlator in~(\ref{e:qtmd}) 
on the renormalization scale is suppressed.
The Wilson line in~(\ref{e:qtmd}) is more complicated than the one for the light-cone
correlators defining the 
GPDs~\cite{Collins:1981uw,Collins:2002kn,Ji:2002aa,Belitsky:2002sm,Boer:2003cm}.
It can be decomposed according to (see also Fig.~\ref{f:3}),
\begin{align} \label{e:wilsontmd}
 &\mathcal{W}_{+\infty}\big(\!-\!\tfrac{1}{2}z;\tfrac{1}{2}z\big)\Big|_{z^+=0^+}\notag\\
 &\ =\big[0^+,-\tfrac{1}{2}z^-,-\tfrac{1}{2}\zT;
  0^+,+\infty^-,-\tfrac{1}{2}\zT\big]\notag\\
 &\spA\times\big[0^+,+\infty^-,-\tfrac{1}{2}\zT;
  0^+,+\infty^-,\tfrac{1}{2}\zT\big]\notag\\
 &\spA\times\big[0^+,+\infty^-,\tfrac{1}{2}\zT;
  0^+,\tfrac{1}{2}z^-,\tfrac{1}{2}\zT\big] \,,
\end{align}
where the future-pointing Wilson lines in~(\ref{e:wilsontmd}) (running along the 
$z^-$ direction in Fig.~\ref{f:3}) are appropriate for defining TMDs in 
SIDIS~\cite{Collins:2002kn}.
In the Drell-Yan process the Wilson lines are necessarily 
past-pointing~\cite{Collins:2002kn},
whereas in hadron-hadron scattering with hadronic final states even more complicated
paths for the Wilson lines can 
arise~\cite{Bomhof:2004aw,Bacchetta:2005rm,Bomhof:2006dp,Bomhof:2006ra}.
Without loss of generality, in this work just the SIDIS case is considered.
If different Wilson lines in (\ref{e:qtmd}) are used, the results of our model 
calculations for the TMDs only differ by calculable factors.

In general the use of lightlike Wilson lines in Eq.~(\ref{e:qtmd}) can lead to 
so-called light-cone divergences~\cite{Collins:1981uw,Collins:2003fm,Gamberg:2006ru}, 
and a refined definition of TMDs is needed in order to avoid this 
problem~\cite{Collins:1981uw,Collins:2003fm,Ji:2004wu,Collins:2004nx,Hautmann:2007uw}.
This issue may make it more difficult to establish model-independent relations 
between GPDs and TMDs.
On the other hand, for the nontrivial relations we are going to study in the context 
of our low order model calculations no light-cone divergence appears, and the use of 
the Wilson line in~(\ref{e:wilsontmd}) is safe. 

The leading twist TMDs are obtained from the correlator in~(\ref{e:qtmd}) by using
the same three matrices $\Gamma$ as in 
Eqs.~(\ref{e:qgpd1})--(\ref{e:qgpd3}),
\begin{align} \label{e:qtmd1}
 &\Phi^{q}(x,\kT;S)=\Phi^{q [\gamma^+]}(x,\kT;S)\notag\\
 &\ =f_1^{q}(x,\kT^{\,2})
  -\frac{\epsilon^{ij}_T k_T^i S_T^j}{M}\,f_{1T}^{\bot q}(x,\kT^{\,2}) \,,
  \displaybreak[0]\\ \label{e:qtmd2}
 &\tilde\Phi^{q}(x,\kT;S)=\Phi^{q [\gamma^+\gamma_5]}(x,\kT;S)\notag\\
 &\ =\lambda\,g_{1L}^{q }(x,\kT^{\,2})
  +\frac{\kT\cdot\ST}{M}\,g_{1T}^{q}(x,\kT^{\,2}) \,,
  \displaybreak[0]\\ \label{e:qtmd3}
 &\Phi_T^{q,\,j}(x,\kT;S)=\Phi^{q [i\sigma^{j+}\gamma_5]}(x,\kT;S)\notag\\
 &\ =-\frac{\epsilon^{ij}_T k_T^i}{M}\,h_1^{\bot q}(x,\kT^{\,2})
  +\frac{\lambda k_T^j}{M}\,h_{1L}^{\bot q}(x,\kT^{\,2})\notag\\
 &\spA+S_T^j\,\bigg(h_{1T}^{q}(x,\kT^{\,2})
  +\frac{\kT^{\,2}}{2M^2}\,h_{1T}^{\bot q}(x,\kT^{\,2})\bigg)\notag\\
 &\spA+\frac{2k_T^j\,\kT\cdot\ST-S_T^j\,\kT^{\,2}}{2M^2}\,h_{1T}^{\bot q}(x,\kT^{\,2}) \,.
\end{align}
We note that a number of other notations exist for some of the quark TMDs (see, e.g., 
Refs.~\cite{Ralston:1979ys,Anselmino:1995vq,Barone:2001sp,Idilbi:2004vb}), 
and that the corresponding TMDs for antiquarks are again given by the same correlation
functions with charge-conjugated fields.

The correlation function for the leading twist gluon TMDs reads
\begin{align} \label{e:gtmd}
 &\Phi^{g[ij]}(x,\kT;S)\notag\\
 &\ =\frac{1}{xP^+}\int\frac{dz^-}{2\pi}\,\frac{d^2\zT}{(2\pi)^2}\,e^{i k\cdot z}\,
  \big<P;S\big|\,
  F^{+j}_a\big(\!-\!\tfrac{1}{2}z\big)\notag\\
 &\spA\times\mathcal{W}_{+\infty,ab}\big(\!-\!\tfrac{1}{2}z;\tfrac{1}{2}z\big)\,
  F^{+i}_b\big(\tfrac{1}{2}z\big)\,\big|P;S\big>\,
  \Big|_{z^+=0^+} \,,
\end{align}
where the path of the Wilson line, now in the adjoint representation, corresponds 
to Eq.~(\ref{e:wilsontmd}). 
It is worthwhile to mention that~(\ref{e:gtmd}) is not the most general gauge 
invariant operator definition for gluon TMDs.
In the most general situation two (different) Wilson lines in the fundamental 
representation can appear (see also, e.g., Ref.~\cite{Bomhof:2006ra}), and only in 
the particular case that these two lines coincide one obtains the definition in 
Eq.~(\ref{e:gtmd}).
However, corresponding to the discussion about more general paths for the Wilson 
line in the quark correlator~(\ref{e:qtmd}), also for the gluon TMDs a more general 
Wilson line in~(\ref{e:gtmd}) would merely change the overall factor of our model 
results and not affect the general conclusions.
For simplicity we therefore restrict ourselves to the definition~(\ref{e:gtmd}).

The twist-2 gluon TMDs are now given through the correlator~(\ref{e:gtmd}) according to
\begin{align} \label{e:gtmd1}
 &\Phi^g(x,\kT;S)=\delta_T^{ij}\,\Phi^{g[ij]}(x,\kT;S)\notag\\
 &\ =f_1^g(x,\kT^{\,2})
  -\frac{\epsilon^{ij}_T k_T^i S_T^j}{M}\,f_{1T}^{\bot g}(x,\kT^{\,2}) \,,
 \displaybreak[0]\\ \label{e:gtmd2}
 &\tilde\Phi^g(x,\kT;S)=i\epsilon_T^{ij}\,\Phi^{g[ij]}(x,\kT;S)\notag\\
 &\ =\lambda\,g_{1L}^g(x,\kT^{\,2})
  +\frac{\kT\cdot\ST}{M}\,g_{1T}^g(x,\kT^{\,2}) \,,
 \displaybreak[0]\\ \label{e:gtmd3}
 &\Phi_T^{g,\,ij}(x,\kT;S)=-\Sop\,\Phi^{g[ij]}(x,\kT;S)\notag\\
 &\ =-\frac{\Sop\,k_T^ik_T^j}{2M^2}\,h_1^{\bot g}(x,\kT^{\,2})
  +\frac{\lambda\,\Sop\,k_T^i\epsilon_T^{jk}k_T^k}{2M^2}\,h_{1L}^{\bot g}(x,\kT^{\,2})\notag\\
 &\spA+\frac{\Sop\,k_T^i\epsilon_T^{jk}S_T^k}{2M}\,\bigg(h_{1T}^g(x,\kT^{\,2})
  +\frac{\kT^{\,2}}{2M^2}\,h_{1T}^{\bot g}(x,\kT^{\,2})\bigg)\notag\\
 &\spA+\frac{\Sop\,k_T^i\epsilon_T^{jk}
  \big(2k_T^k\,\kT\cdot\ST-S_T^k\,\kT^{\,2}\big)}{4M^3}\,h_{1T}^{\bot g}(x,\kT^{\,2}) \,.
\end{align}
Analogous to the GPD case, the definitions of the chiral-even quark and gluon 
TMDs correspond to each other [compare the RHS of~(\ref{e:qtmd1}) and~(\ref{e:gtmd1}), 
as well as the RHS of (\ref{e:qtmd2}) and~(\ref{e:gtmd2})].
In the chiral-odd sector the definitions of the quark and gluon TMDs are (symbolically) 
connected by 
\begin{equation}
  \Phi_T^{g,\,ij} \leftrightarrow \Sop\,\frac{k_T^i}{2M}\,\epsilon_T^{jk}\Phi_T^{q/\bar q,\,k}
\end{equation}
[compare the RHS of~(\ref{e:qtmd3}) and~(\ref{e:gtmd3})].
The gluon TMDs defined in Eqs.~(\ref{e:gtmd1})--(\ref{e:gtmd3}) are related to those 
of Ref.~\cite{Mulders:2000sh} through 
\begin{align}
 f_1^g&=+G \,, &
  h_1^{\bot g}&=+H^\bot \,, \notag\\
 f_{1T}^{\bot g}&=-G_T \,, &
  h_{1T}^g&=-\Big(\Delta H_T-\tfrac{\kT^{\,2}}{2M^2}\Delta H_T^\bot\Big) \,, \notag\\
 g_{1L}^g&=-\Delta G_L \,, &
  h_{1L}^{\bot g}&=-\Delta H_L^\bot \,, \quad\notag\\
 g_{1T}^g&=-\Delta G_T \,, &
  h_{1T}^{\bot g}&=-\Delta H_T^\bot \,.
\end{align}
In the forward limit $f_1^g$ corresponds to the unpolarized gluon distribution, and
$g_{1L}^g$ coincides with the gluon helicity distribution often denoted by $\Delta G$.
Our notation for the gluon TMDs is a natural extension of the nomenclature for  
quark TMDs. 
In the context of our work we prefer a notation with small letters for the TMDs in 
order to avoid confusion with the nomenclature for GPDs.
Eventually, notice that yet another notation for gluon TMDs was proposed in 
Ref.~\cite{Anselmino:2005sh}.

Altogether there exist eight leading twist quark TMDs and eight leading twist gluon
TMDs, which are all real-valued.
Using commutation relations for the parton fields one can derive the symmetry
relations
\begin{equation}
 Y^{\bar q/g}(x,\kT^{\,2})=\pm\,Y^{q/g}(-x,\kT^{\,2}) 
\end{equation}
for the TMDs $Y$, where the \emph{plus} holds for
\begin{equation}
 f_{1T}^{\bot q/g}, \;
 g_{1L}^{q/g}, \;
 h_1^{\bot q}, \;
 h_{1L}^{\bot q}, \;
 h_{1T}^g, \;
 h_{1T}^{\bot g} \,,
\end{equation}
and the \emph{minus} for
\begin{equation}
 f_1^{q/g}, \;
 g_{1T}^{q/g}, \;
 h_1^{\bot g}, \;
 h_{1L}^{\bot g}, \;
 h_{1T}^{q}, \;
 h_{1T}^{\bot q} \,.
\end{equation}
Like in the GPD case it is therefore sufficient to consider only the region
$x>0$.

The TMDs can also be divided into T-even and T-odd distributions.
Whereas on the quark sector two functions are T-odd~($f_{1T}^{\bot q},\;h_1^{\bot q}$), 
there exist four T-odd gluon TMDs
($f_{1T}^{\bot g},\;h_{1L}^{\bot g},\;h_{1T}^{g},\;h_{1T}^{\bot g}$).

The transverse momentum dependent transversity 
distribution of quarks and gluons is given by
\begin{equation}\label{e:trans}
  h_{1T}^{q/g}(x,\kT^{\,2})
  +\frac{\kT^{\,2}}{2M^2}\,h_{1T}^{\bot q/g}(x,\kT^{\,2}) \,.
\end{equation}
In the forward limit, where the parton correlators are integrated upon the transverse
parton momentum, the gluon transversity drops out.
This result is obvious from its prefactor in Eq.~(\ref{e:gtmd3}), 
and ultimately follows from conservation of angular momentum.

With the exception of $f_1$ all TMDs characterize the strength of different
spin-spin and spin-orbit correlations.
The precise form of these correlations is given by the prefactors of the TMDs
in Eqs.~(\ref{e:qtmd1})--(\ref{e:qtmd3}) and~(\ref{e:gtmd1})--(\ref{e:gtmd3}).
To be more specific, the TMD $g_{1L}$ and the transversity distribution in Eq.~(\ref{e:trans})
describe the strength of a correlation 
between a longitudinal/transverse target polarization and a
longitudinal (circular)/transverse (linear) parton polarization. 
By definition the spin-orbit correlations invoke the transverse parton momentum 
and either a polarization of the target~($f_{1T}^\bot$), or of the parton~($h_1^\bot$),
or of both~($g_{1T},\;h_{1L}^\bot,\;h_{1T}^\bot$).
Note that a corresponding discussion on spin-spin and spin-orbit correlations
also applies to the GPDs in impact parameter representation 
(see, e.g, Ref.~\cite{Diehl:2005jf}), where the role of the transverse parton 
momentum in the case of TMDs is played by the impact parameter.
In this context the only difference between TMDs and GPDs lies in the fact that
the spin-orbit correlations accompanied by the functions $g_{1T}$ and $h_{1L}^\bot$
have no counterpart on the GPD side for $\xi=0$ because the corresponding
GPDs $\tilde{E}$ and $\tilde{E}_T$ do not show up as explained after
Eq.~(\ref{e:gpdabbr}) (see also the related discussion in Ref.~\cite{Diehl:2005jf}).

\section{Relations between GPD\lowercase{s} and TMD\lowercase{s}:
model-independent considerations}
\label{s:three}
In this section the current knowledge on attempts to establish model-independent
nontrivial relations between GPDs and TMDs is summarized.
As already mentioned earlier, a special role is played by the impact parameter 
representation of the GPDs, which was used in Ref.~\cite{Burkardt:2003uw} with the aim 
to find a relation between the GPD $E$ and the Sivers function $f_{1T}^\perp$.
In~\cite{Diehl:2005jf} the impact parameter picture was exploited, in particular, to 
write down (model-independent) analogies between chiral-odd quark GPDs and TMDs.
Here the treatment of Ref.~\cite{Diehl:2005jf} is extended to the gluon sector.
It is also argued that it is possible to consider GPDs in momentum instead of
impact parameter space if one wants to have some guidance for possible relations 
between the two types of parton distributions.
In addition, we briefly comment on the representation of parton distributions through 
light-cone wave functions, which also can give some hint on nontrivial relations 
between GPDs and TMDs.

\subsection{Forward parton distributions}
For completeness we first recall the well-known trivial relations between GPDs 
and TMDs.
These relations hold due to the connection between GPDs and TMDs on the one hand
and ordinary forward parton distributions (PDFs) on the other.
For the specific kinematics $\xi = t = 0$ some GPDs are related to the three 
twist-2 PDFs: $f_1(x)$ (unpolarized distribution), $g_1(x)$ (helicity distribution), 
and $h_1(x)$ (transversity distribution)~\cite{Ralston:1979ys}.
The same applies to some TMDs if they are integrated upon the transverse parton 
momentum.
To be specific, in our notation one has
\begin{align} \label{e:trivial1}
 &f_1^{q/g}(x)=\int d^2\kT\,f_1^{q/g}(x,\kT^{\,2})\notag\\
 &\ =H^{q/g}(x,0,0)=\int d^2\bT\,\mathcal{H}^{q/g}(x,\bT^{\,2}) \,,
 \displaybreak[0]\\ \label{e:trivial2}
 &g_1^{q/g}(x)=\int d^2\kT\,g_{1L}^{q/g}(x,\kT^{\,2})\notag\\
 &\ =\tilde{H}^{q/g}(x,0,0)=\int d^2\bT\,\mathcal{\tilde H}^{q/g}(x,\bT^{\,2}) \,,
 \displaybreak[0]\\ \label{e:trivial3}
 &h_1^{q}(x)=\int d^2\kT\,\bigg(h_{1T}^{q}(x,\kT^{\,2})
  +\frac{\kT^{\,2}}{2M^2}\,h_{1T}^{\bot q}(x,\kT^{\,2})\bigg) \notag\\
 &\ = H_T^q(x,0,0) \vphantom{\bigg(} \notag\\
 &\ =\int d^2\bT\,\bigg(\mathcal{H}_T^{q}(x,\bT^{\,2})
  -\frac{\bT^{\,2}}{M^2}\,\Delta_b\mathcal{\tilde H}_T^{q}(x,\bT^{\,2})\bigg) \,.
\end{align}

Concerning the relations between TMDs and PDFs a word of caution is in order.
Already in Sec.~\ref{s:two}\,C we mentioned that in general a refined definition for 
TMDs, containing certain non-lightlike Wilson lines, is needed to avoid possible 
problems with light-cone divergences.
For such a definition the relations between PDFs and TMDs in 
Eqs.~(\ref{e:trivial1})--(\ref{e:trivial3}), in general, turn out to be nontrivial
(see, e.g., Refs.~\cite{Collins:2003fm,Ji:2004wu,Hautmann:2007uw}).
However, provided that one avoids the endpoints $x=0$ and $x=1$, for our low order 
model calculations the definition of TMDs as given 
in~(\ref{e:qtmd}) and~(\ref{e:wilsontmd}) does not lead to any problem and all 
relations~(\ref{e:trivial1})--(\ref{e:trivial3}) are satisfied.

\subsection{Average transverse momentum}
In Ref.~\cite{Burkardt:2003uw} an attempt was made to obtain a nontrivial relation 
between the GPD $E^q$ (in impact parameter representation) and the Sivers function 
$f_{1T}^{\bot q}$.
The object considered there is the average transverse momentum of a quark in a
transversely polarized target.
The main steps of the treatment of~\cite{Burkardt:2003uw} are repeated here.
While in~\cite{Burkardt:2003uw} the light-cone gauge $A^+=0$ was used, we do not work
in a specific gauge.

The average transverse momentum of an unpolarized quark in a transversely polarized
target is defined by 
\begin{align} \label{e:transvmom}
 &\big<k_T^{q,i}(x)\big>_{UT} = \int d^2\kT\,k_T^i\,\Phi^q(x,\kT;S) \notag\\
 &\ =\frac{1}{2}\int d^2\kT\,k_T^i\,\big[\Phi^q(x,\kT;\ST)\!-\!\Phi^q(x,\kT;-\ST)\big] \,, \!\!\!
\end{align}
with the correlator $\Phi^q$ in~(\ref{e:qtmd1}).
The second step on the RHS of~(\ref{e:transvmom}) is justified because obviously
only that part of $\Phi^q$ which depends on the transverse spin gives rise to a 
nonvanishing transverse momentum.
In the following $\big<k_T^{q,i}(x)\big>_{UT}$ will be expressed in terms of 
(twist-3) quark-gluon-quark correlation 
functions~\cite{Burkardt:2003uw,Boer:2003cm,Ma:2003ut}. 
To do so, we first rewrite the second term on the RHS of~(\ref{e:transvmom}) by means
of the parity and time-reversal transformation,
\begin{align}
 &\Phi^q(x,\kT;-\ST)\notag\\
 &\ =\frac{1}{2}\int\frac{dz^-}{2\pi}\,\frac{d^2\zT}{(2\pi)^2}\,e^{i k\cdot z}\,
  \big<P;-\ST\big|\,
  \bar\psi\big(\!-\!\tfrac{1}{2}z\big)\,\gamma^+\notag\\
 &\spA\times\mathcal{W}_{+\infty}\big(\!-\!\tfrac{1}{2}z;\tfrac{1}{2}z\big)\,
  \psi\big(\tfrac{1}{2}z\big)\,\big|P;-\ST\big>\,
  \Big|_{z^+=0^+}\notag\\
 &\ =\frac{1}{2}\int\frac{dz^-}{2\pi}\,\frac{d^2\zT}{(2\pi)^2}\,e^{i k\cdot z}\,
  \big<P;\ST\big|\,
  \bar\psi\big(\!-\!\tfrac{1}{2}z\big)\,\gamma^+\notag\\
 &\spA\times\mathcal{W}_{-\infty}\big(\!-\!\tfrac{1}{2}z;\tfrac{1}{2}z\big)\,
  \psi\big(\tfrac{1}{2}z\big)\,\big|P;\ST\big>\,
  \Big|_{z^+=0^+} \,.
\end{align}
This leads to the intermediate result
\begin{align} \label{e:transvmom1}
 &\big<k_T^{q,i}(x)\big>_{UT}\notag\\
 &\ =\frac{1}{4}\int d^2\kT\,k_T^i\,\int\frac{dz^-}{2\pi}\,\frac{d^2\zT}{(2\pi)^2}\,e^{i k\cdot z}\,
  \big<P;\ST\big|\,\bar\psi\big(\!-\!\tfrac{1}{2}z\big)\,\gamma^+\notag\\
 &\spA\times\,\Big[\mathcal{W}_{+\infty}\big(\!-\!\tfrac{1}{2}z;\tfrac{1}{2}z\big)
  -\mathcal{W}_{-\infty}\big(\!-\!\tfrac{1}{2}z;\tfrac{1}{2}z\big)\Big]\notag\\
 &\spA\times\psi\big(\tfrac{1}{2}z\big)\,\big|P;\ST\big>\,\Big|_{z^+=0^+}
  \notag\\
 &\ =\frac{i}{4}\int\frac{dz^-}{2\pi}\,e^{i k\cdot z}\,\partial_T^i\,
  \big<P;\ST\big|\,\bar\psi\big(\!-\!\tfrac{1}{2}z\big)\,\gamma^+\notag\\
 &\spA\times\,\Big[\mathcal{W}_{+\infty}\big(\!-\!\tfrac{1}{2}z;\tfrac{1}{2}z\big)
  -\mathcal{W}_{-\infty}\big(\!-\!\tfrac{1}{2}z;\tfrac{1}{2}z\big)\Big]\notag\\
 &\spA\times\psi\big(\tfrac{1}{2}z\big)\,\big|P;\ST\big>\,
  \Big|\,\!_{\substack{z^+=0^+\\\zT=\nT}} \,.
\end{align}
In order to perform the second step in~(\ref{e:transvmom1}) we have expressed the factor 
$k_T^i$ through the transverse derivative $\partial_T^i$ acting on the exponential, 
and performed an integration by parts.

The derivative in~(\ref{e:transvmom1}) can now act either on the quark fields or on the 
Wilson lines. 
In the first case though, one gets no contribution to the average transverse momentum, 
since the involved combination of Wilson lines vanishes, 
\begin{equation} \label{e:wilsonvan}
 \Big[\mathcal{W}_{+\infty}\big(\!-\!\tfrac{1}{2}z;\tfrac{1}{2}z\big)-
 \mathcal{W}_{-\infty}\big(\!-\!\tfrac{1}{2}z;\tfrac{1}{2}z\big)\Big]\,
 \Big|\,\!_{\substack{z^+=0^+\\\zT=\nT}}=0 \,.
\end{equation}
The result~(\ref{e:wilsonvan}) is obvious because both Wilson lines are just
running along the light-cone.

On the other hand, if the derivative acts on the Wilson lines, one finds
\begin{align}
 &i\,\partial_T^i\,\Big[\mathcal{W}_{+\infty}\big(\!-\!\tfrac{1}{2}z;\tfrac{1}{2}z\big)-
  \mathcal{W}_{-\infty}\big(\!-\!\tfrac{1}{2}z;\tfrac{1}{2}z\big)\Big]\,
  \Big|\,\!_{\substack{z^+=0^+\\\zT=\nT}}\notag\\
 &\ =g\int dy^-\,\mathcal{W}\big(\!-\!\tfrac{1}{2}z;y\big)\,t_a\,F^{+i}_a\big(y\big)
  \mathcal{W}\big(y;\tfrac{1}{2}z\big)\,
  \Big|\,\!_{\substack{y^+=z^+=0^+\\\yT=\zT=\nT}}\notag\\
 &\ =2\,\mathcal{W}\big(\!-\!\tfrac{1}{2}z;\tfrac{1}{2}z\big)\,
  I^{q,i}\big(\tfrac{1}{2}z\big)\,
  \Big|\,\!_{\substack{z^+=0^+\\\zT=\nT}} \,,
\end{align}
where the paths of the remaining Wilson lines run along the light-cone and
the function $I^{q,i}$ is defined by
\begin{align}
 &I^{q,i}\big(\tfrac{1}{2}z\big)\notag\\
 &=\frac{g}{2}\int dy^-\,\mathcal{W}\big(\tfrac{1}{2}z;y\big)\,
  t_a\,F^{+i}_a\big(y\big)\,\mathcal{W}\big(y;\tfrac{1}{2}z\big)\,
  \Big|\,\!_{\begin{subarray}{l}y^+=z^+\\\yT=\zT\end{subarray}} \,.
\end{align}
Plugging the results together one arrives at the following expression for the average 
transverse momentum,
\begin{align} \label{e:transvmom2}
 &\big<k_T^{q,i}(x)\big>_{UT}\notag\\
 &\ =\frac{1}{2}\int\frac{dz^-}{2\pi}\,e^{i k\cdot z}\,
  \big<P;\ST\big|\,\bar\psi\big(\!-\!\tfrac{1}{2}z\big)\,\gamma^+\notag\\
 &\spA\times\mathcal{W}\big(\!-\!\tfrac{1}{2}z;\tfrac{1}{2}z\big)\,
  I^{q,i}\big(\tfrac{1}{2}z\big)\,\psi\big(\tfrac{1}{2}z\big)\,\big|P;\ST\big>\,
  \Big|\,\!_{\substack{z^+=0^+\\\zT=\nT}} \,. \!\!\!\!
\end{align}
Equation~(\ref{e:transvmom2}) is a representation of the average transverse momentum in terms 
of a specific quark-gluon-quark light-cone 
correlator~\cite{Burkardt:2003uw,Boer:2003cm,Ma:2003ut}. 
Since the gluon field in the three-parton correlator in~(\ref{e:transvmom2}) has 
zero longitudinal momentum one often talks about a soft gluon matrix 
element. 
The reader is referred to~\cite{Efremov:1981sh,Efremov:1984ip,Qiu:1991pp,Qiu:1991wg}
where such (or similar) matrix elements were first discussed in connection with 
transverse SSAs.

To unravel a possible connection between the Sivers effect and the GPD $E^q$, in 
Ref.~\cite{Burkardt:2003uw} the RHS of~(\ref{e:transvmom2}) was transformed to 
the impact parameter space, where it takes the form
\begin{align} \label{e:transvmom3}
 &\big<k_T^{q,i}(x)\big>_{UT}\notag\\
 &\ =\frac{1}{2}\int d^2\bT\,\int\frac{dz^-}{2\pi}\,e^{ixP^+z^-}\,
  \big<P^+,\nT;S\big|\,\bar\psi\big(z_1\big)\,\gamma^+\notag\\
 &\spA\times\mathcal{W}\big(z_1;z_2\big)\,
  I^{q,i}\big(z_2\big)\,\psi\big(z_2\big)\,\big|P^+,\nT;S\big> \,,
\end{align}
with $z_{1/2}$ as given in Eq.~(\ref{e:zdef}).
Comparing the expression in~(\ref{e:transvmom3}) with the correlator~(\ref{e:qimpact})
for the quark GPDs in impact parameter space (for $\Gamma = \gamma^+$) one realizes 
that the only difference is the additional factor $I^{q,i}$ and an integration upon 
the impact parameter $\bT$~\cite{Burkardt:2003uw}.
On the basis of this observation one may hope to find a relation of the type
\begin{align} \label{e:transvmom4}
 &\big<k_T^{q,i}(x)\big>_{UT}
    = \int d^2\kT\,k_T^i\,\Phi^q(x,\kT;S) \notag\\
 &\ \simeq \int d^2\bT\,\mathcal{I}^{q,i}(x,\bT)\,\mathcal{F}^q(x,\bT;S) \,,
\end{align}
where, in rough terms, the function $\mathcal{I}^{q,i}$ incorporates the effect of the 
gluon field in the correlator on the RHS of~(\ref{e:transvmom2}).
We mention that in the second term on the RHS of~(\ref{e:transvmom4})
only the spin-dependent term of ${\mathcal F}^q$ contributes.

Expressed in terms of TMDs and GPDs Eq.~(\ref{e:transvmom4}) reads
\begin{align} \label{e:transvmom5}
 &\big<k_T^{q,i}(x)\big>_{UT}\notag\\
 &\ =-\int d^2\kT\,k_T^i\,\frac{\epsilon_T^{jk} k_T^j S_T^k}{M}\,f_{1T}^{\bot q}(x,\kT^{\,2})\notag\\
 &\ \simeq \int d^2\bT\,\mathcal{I}^{q,i}(x,\bT)\,\frac{\epsilon_T^{jk} b_T^j S_T^k}{M}\,
  \bigg(\mathcal{E}^q(x,\bT^{\,2})\bigg)' \,.
\end{align}
Interestingly, the relation~(\ref{e:transvmom5}) is indeed fulfilled in the context of 
perturbative low order model calculations~\cite{Burkardt:2003je} 
(see also Sec.~\ref{s:four}).
It also provides an intuitive understanding of the origin of the Sivers transverse 
SSA~\cite{Burkardt:2002ks,Burkardt:2003uw}.
However, Eq.~(\ref{e:transvmom5}) does not have the status of a general, model-independent 
result (see also, e.g., Ref.~\cite{Burkardt:2006td}).
The crucial problem lies in the fact that, in general, the average transverse momentum 
$\big<k_T^{q,i}(x)\big>_{UT}$ caused by the Sivers effect cannot be factorized into the 
function $\mathcal{I}^{q,i}$ (called lensing function in~\cite{Burkardt:2003uw}) and the 
distortion of the impact parameter distribution of quarks in a transversely polarized 
target which is determined by $(\mathcal{E}^q)'$.

\subsection{Generalization of relations}
To get further insight into possible relations between GPDs and TMDs, which at least may 
hold in the context of model calculations, we now follow a procedure given in 
Ref.~\cite{Diehl:2005jf}.
The equations defining the GPDs in impact parameter space 
[see Eqs.~(\ref{e:impact1})--(\ref{e:impact4})] on the one hand and the TMDs 
[see Eqs.~(\ref{e:qtmd1})--(\ref{e:qtmd3}) and (\ref{e:gtmd1})--(\ref{e:gtmd3})] on the 
other obviously have a corresponding structure if one interchanges the impact parameter 
$\bT$ and the transverse momentum $\kT$.
Comparing these equations one directly finds out which functions may be related.
However, using this procedure one cannot extract the precise form of the relations.
Note also that the two TMDs $g_{1T}$ and $h_{1L}^\bot$ have no counterpart on the GPD 
side, as already pointed out in Sec.~\ref{s:two}\,C.
In the following we, respectively, talk about relations of first, second, third, and fourth 
type, depending on the number of derivatives of the involved GPDs in impact parameter space.
In the case of quark distributions the results given in this subsection were already presented 
in Ref.~\cite{Diehl:2005jf}. 
At this point one has to keep in mind that, apart from the trivial model-independent
relations (relations of first type), all relations presented in this and the following
subsection so far have only the status of analogies between functions which follow
from obvious analogies in the structures of the GPD and TMD correlators.
Quantitative relations will be discussed in Sec.~\ref{s:four} in connection with
model calculations.

First of all, one finds the following connections by means of the mentioned comparison,
\begin{align} \label{e:type1}
 f_1^{q/g}\leftrightarrow\mathcal{H}^{q/g} \,,
  &\qquad g_{1L}^{q/g}\leftrightarrow\mathcal{\tilde H}^{q/g} \,, \notag\\
 \Big(h_{1T}^{q}+\tfrac{\kT^{\,2}}{2M^2}\,h_{1T}^{\bot q}\Big)
  &\leftrightarrow\Big(\mathcal{H}_T^{q}-\tfrac{\bT^{\,2}}{M^2}\,\Delta_b\mathcal{\tilde H}_T^{q}\Big) \,,
\end{align}
which simply correspond to the trivial relations discussed in Sec.~\ref{s:three}\,A.

Relations of second type contain GPDs with one derivative,
\begin{align} \label{e:type2}
 f_{1T}^{\bot q/g}\leftrightarrow-\Big(\mathcal{E}^{q/g}\Big)' \,,
  &\qquad h_1^{\bot q}\leftrightarrow
  -\Big(\mathcal{E}_T^{q}+2\mathcal{\tilde H}_T^{q}\Big)' \,, \notag\\
 \Big(h_{1T}^g+\tfrac{\kT^{\,2}}{2M^2}\,h_{1T}^{\bot g}\Big)
  &\leftrightarrow -2\Big(\mathcal{H}_T^g-\tfrac{\bT^{\,2}}{M^2}\,\Delta_b\mathcal{\tilde H}_T^g\Big)' \,,
\end{align}
where the first relation in~(\ref{e:type2}) involving $f_{1T}^{\bot q}$ and the derivative of
$\mathcal{E}^q$ corresponds to Eq.~(\ref{e:transvmom5}). 
At this point it is also worthwhile to notice that the computation of the average transverse 
momentum of a transversely polarized quark in an unpolarized target, using the correlator 
in Eq.~(\ref{e:qtmd3}), can be carried out completely analogous to Sec.~\ref{s:three}\,B 
above where the transverse momentum caused by the Sivers effect is considered.
Doing so, one eventually obtains an equation corresponding to~(\ref{e:transvmom5}), with the 
quark Boer-Mulders function $h_1^{\bot q}$ showing up on the TMD side, and the first derivative 
of the linear combination $\mathcal{E}_T^q+2\mathcal{\tilde H}_T^q$ on the GPD side.
On the basis of these considerations one, in particular, also expects the same lensing function 
$\mathcal{I}^{q,i}$ to appear in the analogue of Eq.~(\ref{e:transvmom5}).
This feature indeed emerges in the context of the model calculations presented in 
Sec.~\ref{s:four}.
We note that a corresponding discussion also holds for both relations between gluon GPDs 
and TMDs in Eq.~(\ref{e:type2}). 

Finally, one obtains two relations of third type, containing GPDs with a second 
derivative, 
\begin{equation} \label{e:type3}
 h_{1T}^{\bot q}\leftrightarrow 2\Big(\mathcal{\tilde H}_T^{q}\Big)'' \,,
  \qquad h_1^{\bot g}\leftrightarrow 2\Big(\mathcal{E}_T^g+2\mathcal{\tilde H}_T^g\Big)'' \,,
\end{equation}
and on the gluon sector even one relation where a GPD enters with its third derivative,
\begin{equation} \label{e:type4}
 h_{1T}^{\bot g}\leftrightarrow -4\Big(\mathcal{\tilde H}_T^g\Big)''' \,.
\end{equation}

We emphasize that the number of derivatives of the GPDs in the various relations in a first
place merely distinguishes between the different type of relations, even though in specific 
model calculations one may find relations with exactly the number of derivatives showing
up in~(\ref{e:type1})--(\ref{e:type4}).
For the relations of first, second, and third type in Eqs.~(\ref{e:type1})--(\ref{e:type3})
this works, e.g., in the context of simple spectator models as outlined in Sec.~\ref{s:four} 
below.
We cannot check this feature for the relation of fourth type~(\ref{e:type4}) by our model 
calculations because the respective parton distributions vanish.

\subsection{Relations in momentum space}
The relations presented in the previous subsection were obtained by comparing the 
equations defining the GPDs in the impact parameter representation (for $\xi=0$) with 
those defining the TMDs.
Here we argue that there is actually no need to make the Fourier transform to the
impact parameter space.
Relations corresponding to~(\ref{e:type1})--(\ref{e:type4}) also emerge
if one compares Eqs.~(\ref{e:qtmd1})--(\ref{e:qtmd3}) and~(\ref{e:gtmd1})--(\ref{e:gtmd3})
on the TMD side with the momentum space correlators for GPDs in~(\ref{e:qgpd1})--(\ref{e:qgpd3}) 
and~(\ref{e:ggpd1})--(\ref{e:ggpd3}) evaluated at $\xi=0$.

At the particular kinematical point $\xi=0$ the RHS of Eqs.~(\ref{e:qgpd1})--(\ref{e:qgpd3}) 
and~(\ref{e:ggpd1})--(\ref{e:ggpd3}) can be simplified considerably.
Using the spin vector $S$ introduced in Sec.~\ref{s:two}\,B one finds after
straightforward algebra
\begin{align} \label{e:gpdxi1}
 &F^{q/g}(x,\Delta_T;S) \notag\\
 &\ =H^{q/g}(x,0,\!-\DT^2)
  -\frac{i\epsilon^{ij}_T \Delta_T^i S_T^j}{2M}\, E^{q/g}(x,0,\!-\DT^2) \,, \!\!\!\!\!
  \displaybreak[0]\\ \label{e:gpdxi2} 
 &\tilde{F}^{q/g}(x,\Delta_T;S) \notag\\
 &\ =\lambda\,\tilde{H}^{q/g}(x,0,-\DT^2) \,,
  \displaybreak[0]\\ \label{e:gpdxi3}
 &F_T^{q,\,j}(x,\Delta_T;S) \notag\\
 &\ =-\frac{i\epsilon^{ij}_T \Delta_T^i}{2M}\,
  \bigg(E_T^q(x,0,-\DT^2)+2\tilde{H}_T^q(x,0,-\DT^2)\bigg) \notag\\
 &\spA+S_T^j\,\bigg(H_T^q(x,0,-\DT^2)
  +\frac{\DT^2}{4M^2}\,\tilde{H}_T^q(x,0,-\DT^2)\bigg) \notag\\
 &\spA-\frac{2\Delta_T^j\,\DT\cdot\ST-S_T^j\,\DT^2}{4M^2}\,\tilde{H}_T^q(x,0,-\DT^2) \,,
  \displaybreak[0]\\ \label{e:gpdxi4} 
 &F_T^{g,\,ij}(x,\Delta_T;S) \notag\\
 &\ =\frac{\Sop\,\Delta_T^i\Delta_T^j}{4M^2}\,
  \bigg(E_T^g(x,0,-\DT^2)+2\tilde{H}_T^g(x,0,-\DT^2)\bigg) \notag\\
 &\spA+\frac{i\Sop\,\Delta_T^i\epsilon_T^{jk}S_T^k}{2M}\,\bigg(H_T^g(x,0,-\DT^2) \notag\\
 &\spB+\frac{\DT^2}{4M^2}\,\tilde{H}_T^g(x,0,-\DT^2)\bigg) \notag\\
 &\spA-\frac{i\Sop\,\Delta_T^i\epsilon_T^{jk}
  \big(2\Delta_T^k\,\DT\cdot\ST-S_T^k\,\DT^2\big)}{8M^3} \notag\\
 &\spB\times\tilde{H}_T^g(x,0,-\DT^2) \,.
\end{align}
We repeat that the GPDs $\tilde{E}$ and $\tilde{E}_T$ do not show up in the expressions above 
because of the choice $\xi=0$.

One readily observes that the structure of Eqs.~(\ref{e:gpdxi1})--(\ref{e:gpdxi4}) corresponds 
to the structure of~(\ref{e:qtmd1})--(\ref{e:qtmd3}) and~(\ref{e:gtmd1})--(\ref{e:gtmd3}).
To be more specific, if one replaces $-i\DT/2$ in the prefactors of the former equations by 
the transverse momentum $\kT$ one recovers the various prefactors of the latter.
This strong similarity of the correlators for GPDs (in momentum space) and TMDs also allows 
one, like in Sec.~\ref{s:three}\,C, to write down relations between the two types 
of parton distributions.
In the following we refer again to relations of first, second, third, and fourth type, 
where the relations of first type are given by
\begin{align} \label{e:type1n}
 f_1^{q/g}\leftrightarrow H^{q/g} \,,
  &\qquad g_{1L}^{q/g}\leftrightarrow {\tilde H}^{q/g} \,, \notag\\
 \Big(h_{1T}^{q}+\tfrac{\kT^{\,2}}{2M^2}\,h_{1T}^{\bot q}\Big)
  &\leftrightarrow \Big(H_T^q+\tfrac{\DT^2}{4M^2}\,{\tilde H}_T^q\Big) \,.
\end{align}
The relations of second type read
\begin{align} \label{e:type2n}
 f_{1T}^{\bot q/g}\leftrightarrow-\Big(E^{q/g}\Big) \,,
  &\qquad h_1^{\bot q}\leftrightarrow
  -\Big(E_T^{q}+2{\tilde H}_T^{q}\Big) \,, \notag\\
 \Big(h_{1T}^g+\tfrac{\kT^{\,2}}{2M^2}\,h_{1T}^{\bot g}\Big)
  &\leftrightarrow -2\Big(H_T^g+\tfrac{\DT^2}{4M^2}\,{\tilde H}_T^g\Big) \,,
\end{align}
and one finds 
\begin{equation} \label{e:type3n}
 h_{1T}^{\bot q}\leftrightarrow 2\Big({\tilde H}_T^{q}\Big) \,,
  \qquad h_1^{\bot g}\leftrightarrow 2\Big(E_T^g+2{\tilde H}_T^g\Big) \,,
\end{equation}
for the relations of third type, as well as
\begin{equation} \label{e:type4n}
 h_{1T}^{\bot g}\leftrightarrow -4\Big({\tilde H}_T^g\Big) \,.
\end{equation}
for the one relation of fourth type.
One may wonder why we distinguish here between different types of relations.
The reason is simply that, in the defining equations, the prefactors of the GPDs and
TMDs carry different powers of $\DT$ and $\bT$, respectively.
If there appears no $\DT$ or $\bT$ in the prefactor we talk about a relation of first type, 
if they appear linearly we talk about a relation of second type and so on.
In connection with this discussion it may be worthwhile to mention that the number of 
derivatives of the GPDs in impact parameter representation matches with the power 
of $\DT$ in the corresponding prefactors in Eqs.~(\ref{e:gpdxi1})--(\ref{e:gpdxi4}).
This correspondence just reflects a property of the Fourier transform from the 
momentum space to the impact parameter space.

Observe now the close analogy between the relations in~(\ref{e:type1n})--(\ref{e:type4n}) 
and those in~(\ref{e:type1})--(\ref{e:type4}).
In the context of our model calculations in Sec.~\ref{s:four} we will consider relations 
between TMDs and GPDs in both impact parameter space as well as momentum space.

Eventually, let us briefly comment on nontrivial relations between GPDs and TMDs that 
can be obtained by looking at a light-cone wave function representation of these 
objects.
This issue was, in particular, discussed in connection with a possible relation
between the quark Sivers function and the GPD $E^q$ 
(see, e.g., Refs.~\cite{Brodsky:2002cx,Brodsky:2006ha,Brodsky:2006hj,Lu:2006kt}).
One finds that for both parton distributions the same light-cone wave functions 
appear, which hints at a close relation between $f_{1T}^{\bot q}$ and $E^q$.
However, it turns out that in the case of the Sivers function one has to augment the 
involved wave functions by a phase factor not present in the case of 
$E^q$~\cite{Brodsky:2006ha}.
This phase spoils a model-independent one-to-one correspondence between $f_{1T}^{\bot}$ 
and $E^q$.
So far only in the context of simple models the Sivers function can be represented
through real wave functions multiplied by some additional 
factor~\cite{Brodsky:2002cx,Lu:2006kt},
allowing one to establish a direct connection between the two distributions.

Moreover, in general (in full QCD) the representation of parton distributions in 
terms of light-cone wave functions may be problematic. 
This feature was already observed in the case of the unpolarized parton 
distribution $f_1^q(x)$ in Ref.~\cite{Brodsky:2002ue}.

\section{Relations between GPD\lowercase{s} and TMD\lowercase{s}:
results of model calculations}
 \label{s:four}
While the previous Sec.~\ref{s:three} is devoted to model-independent considerations 
on possible relations between GPDs and TMDs, here we discuss the relations in the 
context of specific model results.
To do so, two spectator models are studied: first, the scalar diquark spectator model of 
the nucleon (see, e.g., Ref.~\cite{Brodsky:2002cx}); second, a quark target model 
treated in perturbative QCD.
Some details concerning these models as well as the full list of GPDs and TMDs, computed 
to lowest nontrivial order in perturbation theory, can be found in the appendices.

\subsection{Relations of first type}
We start this discussion with the relations of first type in 
Eqs.~(\ref{e:trivial1})--(\ref{e:trivial3}).
As an example consider the TMD $f_1^q$ in~(\ref{e:sdmtmd1}) and the GPD $H^q$ 
in~(\ref{e:sdmgpd1}) in the scalar diquark model.
Using these results one readily verifies that
\begin{align}
 &f_1^q(x)=\int d^2\kT\,f_1^q(x,\kT^{\,2})\notag\\
 &\ =\frac{g^2(1-x)}{2(2\pi)^3}\,\int d^2\kT\,
  \frac{\kT^{\,2}+(m_q+xM)^2}
  {\big[\kT^{\,2}+\tilde M^2(x)\big]\,\!^2}\notag\\
 &\ =H^q(x,0,0)=\int d^2\bT\,\mathcal{H}^q(x,\bT^{\,2}) \,,
\end{align}
i.e., that the relation~(\ref{e:trivial1}) holds.
The unpolarized distributions of scalar diquarks and of quarks and gluons in the
quark target model satisfy Eq.~(\ref{e:trivial1}) as well.
Moreover, the results in both models obey the relations~(\ref{e:trivial2}) 
and~(\ref{e:trivial3}).
Of course the model results have to satisfy these model-independent relations.
Therefore, the comparison discussed here merely serves as a consistency check 
of the calculation.

\subsection{Relations of second type}
In order to study the relations of second type in Eq.~(\ref{e:type2}) in the context of 
our model calculations we first consider the average transverse momentum of a parton 
caused by the Sivers effect (see Sec.~\ref{s:three}\,B).
Details of the calculation are given for a quark in the scalar spectator model.
It turns out that the transverse momentum can indeed be expressed according to 
Eq.~(\ref{e:transvmom5}) in terms of $(\mathcal{E}^q)'$.  
The reader is referred to~\cite{Burkardt:2003je} where the connection 
in~(\ref{e:transvmom5}) in the context of the spectator model was presented for the 
first time.

By definition the average transverse momentum of an unpolarized quark in a transversely
polarized target is given by [see also~(\ref{e:transvmom5})]
\begin{align} \label{e:transsiv1}
 &\big<k_T^{q,i}(x)\big>_{UT}\notag\\
 &\ =-\int d^2\kT\,k_T^i\,\frac{\epsilon_T^{jk} k_T^j S_T^k}{M}\,f_{1T}^{\bot q}(x,\kT^{\,2})\notag\\
 &\ =-\frac{g^2e_qe_s(1-x)(m_q+xM)}{4(2\pi)^4}\,\int d^2\kT\notag\\
 &\spA\times\frac{k_T^i\,\epsilon_T^{jk} k_T^j S_T^k}
  {\kT^{\,2}\,\big[\kT^{\,2}+\tilde M^2(x)\big]}\,
  \ln\bigg(\frac{\kT^{\,2}+\tilde M^2(x)}{\tilde M^2(x)}\bigg)\notag\\
 &\ =\frac{g^2e_qe_s(1-x)(m_q+xM)}{2(2\pi)^3}\,
  \int\frac{d^2\lT}{(2\pi)^2}\,\frac{l_T^i}{\lT^{\,2}}\,\int d^2\kT\notag\\
 &\spA\times\frac{\epsilon_T^{jk} k_T^j S_T^k}
  {\big[\kT^{\,2}+\tilde M^2(x)\big]\,\big[\big(\kT+\lT\big)\,\!^2+\tilde M^2(x)\big]} \,,
\end{align}
where we have inserted the result for the Sivers function from Eq.~(\ref{e:sdmtmd2}) and used 
in the last step that
\begin{align}
 &\int\frac{d^2\lT}{(2\pi)^2}\,\frac{l_T^i}
  {\lT^{\,2}\,\big[\big(\kT+\lT\big)\,\!^2+\tilde M^2(x)\big]}\notag\\
 &\ =-\frac{k_T^i}{4\pi\,\kT^{\,2}}\,
  \ln\bigg(\frac{\kT^{\,2}+\tilde M^2(x)}{\tilde M^2(x)}\bigg) \,.
\end{align}
If one now replaces the integration variable $\kT$ according to $\kT\rightarrow-\kT-\lT$ one 
gets
\begin{align} \label{e:transsiv2}
 &\!\!\big<k_T^{q,i}(x)\big>_{UT}\notag\\
 &\!\!\ =-\frac{g^2e_qe_s(1-x)(m_q+xM)}{4(2\pi)^3}\,
  \int\frac{d^2\lT}{(2\pi)^2}\,\frac{l_T^i}{\lT^{\,2}}\,\int d^2\kT \notag\\
 &\!\!\spA\times\frac{\epsilon_T^{jk} l_T^j S_T^k}
  {\big[\kT^{\,2}+\tilde M^2(x)\big]\,\big[\big(\kT+\lT\big)\,\!^2+\tilde M^2(x)\big]}\notag\\
 &\!\!\ =-\frac{e_qe_s}{4}\int\!\frac{d^2\lT}{(2\pi)^2}\,\frac{l_T^i}{\lT^{\,2}}
  \,\frac{\epsilon_T^{jk} l_T^j S_T^k}{(1-x)M}\,E^q\Big(x,0,\!-\tfrac{\lT^{\,2}}{(1-x)^2}\Big) \,, \!\!
\end{align}
where we took for $E^q$ the result in Eq.~(\ref{e:sdmgpd2}).
In the next step the transformation to the impact parameter space is performed
which leads to 
\begin{align} \label{e:transsiv3}  
 &\big<k_T^{q,i}(x)\big>_{UT}\notag\\
 &\ =-\frac{e_qe_s}{4}\,\int\frac{d^2\lT}{(2\pi)^2}\,\frac{l_T^i}{\lT^{\,2}} \notag\\
 &\spA\times\frac{\epsilon_T^{jk} l_T^j S_T^k}{(1-x)M}\,
  \int d^2\bT\,e^{i\tfrac{\lT\cdot\bT}{1-x}}\,\mathcal{E}^q(x,\bT^{\,2}) \notag\\
 &\ =\int d^2\bT\,\frac{ie_qe_s}{2}\,
  \int\frac{d^2\lT}{(2\pi)^2}\,e^{-i\tfrac{\lT\cdot\bT}{1-x}}\,\frac{l_T^i}{\lT^{\,2}}\notag\\
 &\spA\times\frac{\epsilon_T^{jk} b_T^j S_T^k}{M}\,\bigg(\mathcal{E}^q(x,\bT^{\,2})\bigg)'\notag\\
 &\ =\int d^2\bT\,\mathcal{I}_{\textrm{SDM}}^{q,i}(x,\bT)\,\frac{\epsilon_T^{jk} b_T^j S_T^k}{M}\,
  \bigg(\mathcal{E}^q(x,\bT^{\,2})\bigg)' \,.
\end{align}
In~(\ref{e:transsiv3}) integration by parts is used in order to obtain a representation 
of the transverse quark momentum that contains the derivative of ${\mathcal E}^q$.
Equation~(\ref{e:transsiv3}) coincides with the relation in~(\ref{e:transvmom5}) between the 
Sivers function and the GPD $E^q$, where the so-called lensing 
function~\cite{Burkardt:2003uw} $\mathcal{I}_{\textrm{SDM}}^{q,i}$ is given by
\begin{align} \label{e:lenssdm}
&\mathcal{I}_{\textrm{SDM}}^{q,i}(x,\bT)=\frac{ie_qe_s}{2}\,
 \int\frac{d^2\lT}{(2\pi)^2}\,e^{-i\tfrac{\lT\cdot\bT}{1-x}}\,\frac{l_T^i}{\lT^{\,2}} \notag\\
&\ =\frac{e_qe_s}{4\pi}\frac{(1-x)\,b_T^i}{\bT^{\,2}} \,.
\end{align}
The calculation in~(\ref{e:transsiv1})--(\ref{e:transsiv3}) also goes through step by 
step if one considers the average transverse momentum of diquarks in the diquark spectator 
model, using the Sivers function $f_{1T}^{\bot s}$ from~(\ref{e:sdmtmd10}) and the GPD $E^s$ 
from~(\ref{e:sdmgpd8}).
One finds that the lensing functions for quarks and diquarks in that model are the same,
\begin{equation}
 \mathcal{I}_{\textrm{SDM}}^{s,i}(x,\bT)=\mathcal{I}_{\textrm{SDM}}^{q,i}(x,\bT) \,.
\end{equation}
Moreover, the transverse momentum $\big<k_T^i(x)\big>_{UT}$ can also be computed
for quarks and gluons in the quark target model.
On the basis of the respective results given in Appendix~\ref{app:b} one obtains in that case 
the lensing functions
\begin{equation} \label{e:lensqtm}
 \mathcal{I}_{\textrm{QTM}}^{q,i}(x,\bT)=\mathcal{I}_{\textrm{QTM}}^{g,i}(x,\bT)=
 -\frac{3g^2}{8\pi}\frac{(1-x)\,b_T^i}{\bT^{\,2}} \,.
\end{equation}
Therefore, all results in the two spectator models satisfy the relation between the
Sivers function and the GPD $E$ in~(\ref{e:type2}), where the specific form of the
relation is given in Eq.~(\ref{e:transvmom5}).
In the context of the model calculations the lensing function does not depend on the 
parton type but merely on the model, which means in other words that it just depends 
on the target.
Note that in both models the lensing function has the same overall sign and is 
negative.

At this point we would like to add some discussion on the intuitive picture of the 
physical origin of the Sivers effect presented in 
Refs.~\cite{Burkardt:2002ks,Burkardt:2003uw}.
The starting point is the observation that the impact parameter distribution 
of unpolarized quarks in a transversely polarized target is 
distorted~\cite{Burkardt:2002hr} (see also the discussion in Sec.~\ref{s:two}\,B),
where the strength of the distortion is determined by the derivative 
of $\mathcal{E}^q$.
A suitable measure for the distortion effect is the flavor dipole moment 
defined by~\cite{Burkardt:2002ks}
\begin{align} \label{e:dipole}
& d^{q,i}=\int dx \int d^2\bT \, b_T^i \, \mathcal{F}^q(x,\bT;S) \notag\\ 
&\ = - \frac{\epsilon_T^{ij}S_T^j}{2M}\int dx \, E^q(x,0,0) 
   = - \frac{\epsilon_T^{ij}S_T^j}{2M} \kappa^q \,,
\end{align}
with the correlator $\mathcal{F}^q$ taken from Eq.~(\ref{e:impact1}).
The dipole moment in~(\ref{e:dipole}) is determined by the contribution $\kappa^q$ 
of the respective quark flavor to the anomalous magnetic moment of the target.
Up to this point the considerations are model-independent.

Using $SU(2)$-flavor symmetry and neglecting in a first approximation the 
contribution from all other partons to the anomalous magnetic moment of the proton 
and the neutron one obtains
\begin{equation} \label{e:amm}
\kappa^{u/p} = \kappa^{d/n} \approx 1.7\,,\qquad
\kappa^{d/p} = \kappa^{u/n} \approx -2.0 \,.
\end{equation}
On the basis of these numbers one finds dipole moments of the order $0.2\,\textrm{fm}$
for the light quark flavors in the nucleon.
This value is quite significant in comparison to the size of 
the nucleon~\cite{Burkardt:2002ks}.

It is now natural to speculate that this large distortion should have an observable 
effect.
In fact, in Refs.~\cite{Burkardt:2002ks,Burkardt:2003uw} it was conjectured that the 
Sivers effect observed in semi-inclusive reactions is intimately related to the
distortion of the quark distributions in a transversely polarized target.
Though quite plausible on a first look, such a connection in general is actually not 
obvious, because impact parameter dependent GPDs \emph{a priori} do not appear in the
QCD-factorization formulas of semi-inclusive processes.
At the moment the only thing known for sure is the following: 
if one describes the Sivers effect (in, e.g., SIDIS) in the framework of a spectator 
model (to lowest nontrivial order), the GPD $\mathcal{E}^q$ enters because of its 
relation to the Sivers function discussed above.
In other words, so far the relation between the distortion of the correlator 
$\mathcal{F}^q$ in impact parameter space and the Sivers effect is only established 
in simple spectator models (see also Ref.~\cite{Burkardt:2006td}).

In the spectator models studied in Ref.~\cite{Burkardt:2003je} and in the present work, 
the Sivers effect, quantified by the average transverse momentum, factorizes into the 
distortion effect times the lensing function according to Eq.~(\ref{e:transsiv3}). 
In SIDIS for instance the lensing function describes the influence of the final 
state interaction of the struck parton.
It is worthwhile to notice that the lensing functions in 
Eqs.~(\ref{e:lenssdm}) and~(\ref{e:lensqtm}) only depend on the variable 
$\vec{c}_T=\bT/(1-x)$ representing the transverse distance between the active parton and 
the spectator parton.
In Ref.~\cite{Burkardt:2003je} it was shown that the lensing function in~(\ref{e:lenssdm}) 
is exactly the net transverse momentum which the active quark acquires due to its 
Coulomb interaction with the spectator diquark, if it moves from the position 
$(c^1,c^2,0)$ to the position $(c^1,c^2,\infty)$.
Therefore, $\big<k_T^{q,i}(x)\big>_{UT}$ in~(\ref{e:transsiv3}) is nothing but this net 
transverse quark momentum convoluted with the transverse position 
distribution~(\ref{e:impact1}) of unpolarized quarks inside a transversely polarized 
target.

What the picture for the Sivers effect given in 
Refs.~\cite{Burkardt:2002ks,Burkardt:2003uw,Burkardt:2003je} predicts is twofold:
first, the Sivers effect for up and down quarks in the nucleon should have opposite 
sign.
To arrive at this conclusion one has to make use of the (model-dependent) factorized 
form of the average transverse momentum in~(\ref{e:transsiv3}), and of the fact that 
the distortion of the distribution $\mathcal{F}^q$ in Eq.~(\ref{e:impact1}) 
(in a model-independent way) is given by the anomalous magnetic moment $\kappa^q$.
The opposite sign then follows from the phenomenological numbers in~(\ref{e:amm}). 
At this point it should be noticed that the different signs are also the outcome 
of an exact, model-independent analysis of the Sivers function in the limit of a 
large number of colors $N_c$~\cite{Pobylitsa:2003ty}.
The second prediction concerns the absolute sign of the Sivers function.
The negative sign of the lensing function~(\ref{e:lenssdm}), reflecting an attractive 
final state interaction of the struck quark, implies for instance 
$f_{1T}^{\bot u/p} < 0$~\cite{Burkardt:2003uw}.
It is interesting that the existing 
analyses~\cite{Efremov:2004tp,Anselmino:2005nn,Anselmino:2005ea,Vogelsang:2005cs,Collins:2005ie,Anselmino:2005an} of the data on the Sivers effect in 
SIDIS~\cite{Airapetian:2004tw,Diefenthaler:2005gx,Alexakhin:2005iw,Ageev:2006da} 
are in accordance with both predictions.

Before proceeding to the Boer-Mulders function for quarks we briefly address the 
gluon Sivers effect in the nucleon.
One can show in a model-independent way that the gluon Sivers function is suppressed 
compared to the Sivers effect of quarks~\cite{Efremov:2004tp}.
This result follows from the Burkardt sum rule for the Sivers 
function~\cite{Burkardt:2003yg,Burkardt:2004ur}, stating that the average transverse momentum of 
unpolarized partons in a transversely polarized target vanishes when summing over 
all partons, and from the large-$N_c$ result according to which the Sivers function 
of the light quarks is equal in magnitude but opposite in sign~\cite{Pobylitsa:2003ty}.
The suppression of the gluon Sivers effect has also been confirmed by recent 
phenomenological analyses~\cite{Anselmino:2006yq,Brodsky:2006ha}.

Now we would like to discuss the relation of second type in Eq.~(\ref{e:type2}) for 
the Boer-Mulders function $h_1^{\bot q}$.
A measure of the Boer-Mulders effect is the average transverse momentum
$\big<k_T^{q,i}(x)\big>_{TU}^j$ of a transversely polarized quark (with polarization
along $j$-direction) in an unpolarized nucleon, which is given by the correlator 
$\Phi_T^{q,\,j}$ in Eq.~(\ref{e:qtmd3}).
A calculation completely analogous to~(\ref{e:transsiv1})--(\ref{e:transsiv3}) in 
the scalar diquark model yields
\begin{align} \label{e:transbm}
 &\big<k_T^{q,i}(x)\big>_{TU}^j \notag\\
 &\ =\frac{1}{2}\int d^2\kT\,k_T^i\,
     \big[\Phi_T^{q,\,j}(x,\kT;S)+\Phi_T^{q,\,j}(x,\kT;-S)\big] \notag\\
 &\ = -\int d^2\kT\,k_T^i\,\frac{\epsilon_T^{kj} k_T^k}{M}\,h_{1}^{\bot q}(x,\kT^{\,2})\notag\\
 &\ =\int d^2\bT\,\mathcal{I}_{\textrm{SDM}}^{q,i}(x,\bT)\,\frac{\epsilon_T^{kj} b_T^k}{M} \notag\\
 &\ \spA\times\bigg(\mathcal{E}_T^q(x,\bT^{\,2})+2\tilde{\mathcal{H}}_T^q(x,\bT^{\,2})\bigg)' \,.
\end{align}
In order to perform the last step in~(\ref{e:transbm}) we used the result for
$h_1^{\bot q}$ from Eq.~(\ref{e:sdmtmd5}) and the result for $E_T^q+2\tilde{H}_T^q$ from 
Eqs.~(\ref{e:sdmgpd5}) and~(\ref{e:sdmgpd6}).
The lensing function showing up in~(\ref{e:transbm}) coincides with the one in 
Eq.~(\ref{e:lenssdm}) for the quark Sivers effect, which means that it is not only 
independent of the parton type but also of its polarization.
This feature is actually not surprising if one keeps in mind that the lensing function
arises from the high-energy (eikonalized) final state interaction of the quark.
Such an eikonalized interaction is not sensitive to the polarization of the quark.
The relation in Eq.~(\ref{e:transbm}) also holds for the results of the respective 
parton distributions obtained in the quark target model, provided that one uses the 
lensing function in~(\ref{e:lensqtm}).

According to Eq.~(\ref{e:transbm}) the (model-dependent) physical picture of the 
Boer-Mulders effect is completely analogous to that of the Sivers effect discussed 
above~\cite{Burkardt:2005hp}.
The magnitude of the Boer-Mulders function is proportional to the distortion of the
impact parameter distribution of transversely polarized quarks in an unpolarized target 
in~(\ref{e:impact3}).
This distortion is given by the derivative of $\mathcal{E}_T^q+2\tilde{\mathcal{H}}_T^q$,
where in the spirit of Eq.~(\ref{e:dipole}) the magnitude of the distortion can be 
quantified in a model-independent way through the object~\cite{Burkardt:2005hp}
\begin{equation}
\kappa_T^q = \int dx \, \Big(E_T^{q}(x,0,0)+2{\tilde H}_T^{q}(x,0,0)\Big) \,.
\end{equation}
While the anomalous magnetic moment $\kappa^q$ is known from experiment [in combination
with reasonable model assumptions, see~(\ref{e:amm})], so far no experimental 
information on $\kappa_T^q$ is available.
However, results on $\kappa_T^q$ were obtained in the framework of a constituent quark 
model~\cite{Pasquini:2005dk} and in lattice QCD~\cite{Gockeler:2006zu}. 
Both studies indicate that in nature $\kappa_T^q$ of the nucleon may be as large or 
perhaps even larger than $\kappa^q$, and that it has the same (positive) sign for the 
two light quarks. 
Analogous to the discussion for the Sivers function, these results for $\kappa_T^q$, 
together with the attractive final state interaction of the quark encoded in the 
lensing function, imply that both $h_1^{\bot u/p}$ and $h_1^{\bot d/p}$ are sizeable 
and negative~\cite{Burkardt:2005hp,Burkardt:2006td}.   
We note in passing that model-independent large-$N_c$ considerations also predict the 
same sign for the Boer-Mulders function of the light quarks in the 
nucleon~\cite{Pobylitsa:2003ty}.

Eventually, we consider the relation of second type in~(\ref{e:type2}) for the 
chiral-odd gluon GPDs and TMDs.
Analogous to the above discussion for the Sivers and Boer-Mulders effect we now study 
the average transverse momentum of a linearly polarized gluon in a transversely 
polarized target.
By definition this object is given by the correlator $\Phi_T^{g,\,ij}$ in 
Eq.~(\ref{e:gtmd3}).
In the framework of the quark target model one finds
\begin{align} \label{e:transgluon}
 &\big<k_T^{g,i}(x)\big>_{LT}^{jk} \notag\\
 &\ =\frac{1}{2}\int d^2\kT\,k_T^i\,
     \big[\Phi_T^{g,\,jk}(x,\kT;\ST)-\Phi_T^{g,\,jk}(x,\kT;-\ST)\big] \notag\\
 &\ = \int d^2\kT\,k_T^i\,\frac{\Sop\,k_T^j\epsilon_T^{kl} S_T^l}{2M} \notag\\
 &\ \spA\times\bigg(h_{1T}^{g}(x,\kT^{\,2})+\frac{\kT^{\,2}}{2M^2}\,
         h_{1T}^{\bot g}(x,\kT^{\,2})\bigg)\notag\\
 &\ =-\int d^2\bT\,\mathcal{I}_{\textrm{QTM}}^{g,i}(x,\bT)\,
         \frac{\Sop\,b_T^j\epsilon_T^{kl} S_T^l}{M} \notag\\
 &\ \spA\times\bigg(\mathcal{H}_T^g(x,\bT^{\,2})-\frac{\bT^{\,2}}{M^2}\,
         \Delta_b\tilde{\mathcal{H}}_T^g(x,\bT^{\,2})\bigg)' \,,
\end{align}
where the two TMDs are taken from Eqs.~(\ref{e:qtmtmd15}) and~(\ref{e:qtmtmd16}), 
and the GPDs from Eqs.~(\ref{e:qtmgpd10}) and~(\ref{e:qtmgpd12}).
We note that the last line in~(\ref{e:gtmd3}) (term proportional to $h_{1T}^{\bot g}$)
does not contribute to the average transverse momentum in~(\ref{e:transgluon}), which 
is a model-independent result.
In Eq.~(\ref{e:transgluon}) the same lensing function as in~(\ref{e:lensqtm}) appears again.
Unlike the average transverse momentum caused by the Sivers or Boer-Mulders
effect, at present nothing is known about the magnitude of the gluon transverse 
momentum in Eq.~(\ref{e:transgluon}).

To summarize, all our spectator model results obey relations as indicated 
in~(\ref{e:type2}).
The specific form of the relations is given in 
Eqs.~(\ref{e:transsiv3}),~(\ref{e:transbm}), and~(\ref{e:transgluon}) and is of exactly 
the same structure for all three relations between GPDs and TMDs in~(\ref{e:type2}).

\subsection{Relations in momentum space}
In the previous subsection an important role was played by the impact parameter 
representation of the GPDs.
Now we want to present relations between TMDs and GPDs in momentum space, where we
again focus on the relations of second type [see the relations indicated 
in Eq.~(\ref{e:type2n})].

For the following considerations it is convenient to introduce moments of the 
GPDs $X$,
\begin{align} \label{e:momgpd}
&X^{(n)}(x) \notag\\
&\ =\frac{1}{2M^2}\int d^2\DT\,\bigg(\frac{\DT^2}{2M^2}\bigg)^{n-1}\,
 X(x,0,-\tfrac{\DT^2}{(1-x)^2}) \,,
\end{align}
and moments of the TMDs $Y$,
\begin{equation} \label{e:momtmd}
 Y^{(n)}(x)=\int d^2\kT\,\bigg(\frac{\kT^{\,2}}{2M^2}\bigg)^n\,Y(x,\kT^{\,2}) \,.
\end{equation}
The relations discussed below are relations between the moments $X^{(n)}$ of the GPDs 
and the moments $Y^{(n)}$ of the TMDs.

On the basis of the diquark model results for $f_{1T}^{\bot q}$ in~(\ref{e:sdmtmd2}) 
and for $E^q$ in~(\ref{e:sdmgpd2}) one finds by explicit calculation
\begin{align} \label{e:type2mom}
&f_{1T}^{\bot q\,(n)}(x) \notag\\
&\ =-\frac{g^2e_qe_s(1-x)}{16(2\pi)^2}\,\frac{(m_q+xM)\tilde M^{2n-2}(x)\,H_{-n}}{2^nM^{2n-1}\,\sin(n\pi)}\notag\\
&\ =-\frac{e_qe_s}{2(2\pi)^2\,(1-x)}\,\frac{H_{-n}\,\Gamma(2-2n)}{\Gamma^2(1-n)}
 \,E^{q\,(n)}(x) \,.
\end{align}
The relation~(\ref{e:type2mom}) holds for $0\leq n \leq 1$, i.e., $n$ is not necessarily 
an integer.
In~(\ref{e:type2mom}) the $\Gamma$-function enters as well as $H_n$, the
analytic continuation of the harmonic numbers for noninteger $n$ given by
\begin{equation}
 H_n=\sum_{i=1}^n\frac{1}{i}
 =\int_0^1dx\,\frac{1-x^n}{1-x}\,.
\end{equation}

Observe the relative factor $(1-x)$ between the moments of the Sivers function and 
those of $E^q$ in Eq.~(\ref{e:type2mom}).
The same relative factor was also found for instance in Ref.~\cite{Brodsky:2006hj} 
by a model-independent investigation on the basis of light-cone wave functions.

Relation~(\ref{e:type2mom}) also holds for the respective functions in the quark 
target model, provided that one replaces the quark and diquark couplings according to 
\begin{equation} \label{e:couprep}
e_qe_s \rightarrow -\tfrac{3}{2}g^2 \,.
\end{equation}
This difference between the two models corresponds exactly to the difference between 
the lensing functions in Eqs.~(\ref{e:lenssdm}) and~(\ref{e:lensqtm}).

Our results in both the scalar diquark model and the quark target model obey all 
relations given in~(\ref{e:type2n}). 
The specific form of the relations between the different parton distributions 
coincides with Eq.~(\ref{e:type2mom}).

In a next step~(\ref{e:type2mom}) is evaluated for three particular values of $n$,
where for the most interesting cases one obtains
\begin{align} \label{e:type2a}
 f_{1T}^{\bot q\,(0)}(x) &= \frac{\pi e_qe_s}{48(1-x)}\,E^q(x,0,0) \,, \\
 \label{e:type2b}
 f_{1T}^{\bot q\,(1/2)}(x) &=\frac{2\ln2\,e_qe_s}{(2\pi)^3\,(1-x)}\,E^{q\,(1/2)}(x) \,,\\
 \label{e:type2c}
 f_{1T}^{\bot q\,(1)}(x) &= \frac{e_qe_s}{4(2\pi)^2\,(1-x)}\,E^{q\,(1)}(x) \,.
\end{align}
We would like to add some discussion on these results by starting with the 
relation in Eq.~(\ref{e:type2c}).
This representation of the relation between the Sivers function and the GPD $E^q$ was 
in principle already given in Ref.~\cite{Burkardt:2003je}, and is equivalent
to Eq.~(\ref{e:transsiv2}) of the present work.
This can be readily seen if one keeps in mind that
\begin{equation}
\int d^2\lT \, l_T^i l_T^j \, A(\lT^{\,2}) =
\tfrac{1}{2}\delta_T^{ij} \int d^2\lT \, \lT^{\,2} \, A(\lT^{\,2}) \,,
\end{equation}
which holds for an arbitrary function $A(\lT^{\,2})$.
The relation~(\ref{e:type2c}) looks simpler than the one in~(\ref{e:transsiv3}) 
containing the GPD in impact parameter representation.
On the other hand, Eq.~(\ref{e:transsiv3}) provides an intuitive physical interpretation 
of the Sivers effect (see Ref.~\cite{Burkardt:2003je} and the discussion in 
Sec.~\ref{s:four}\,B).

The relation in Eq.~(\ref{e:type2a}) was already recently discovered in 
Ref.~\cite{Lu:2006kt}.
Starting from~(\ref{e:type2mom}) one has to make use of the result
\begin{equation} \label{e:modelid}
 \lim_{n\rightarrow0}\,\frac{H_{-n}\,\Gamma(2-2n)}{\Gamma^2(1-n)}\,E^{q\,(n)}(x)
 =-\frac{\pi^3}{6}\,E^q(x,0,0) 
\end{equation} 
in order to arrive at the relation~(\ref{e:type2a}).
The identity in~(\ref{e:modelid}) is valid for the result of $E^q$ in the simple spectator 
models considered in the present work but may not hold in general.

Eventually, the case $n=\tfrac{1}{2}$ in~(\ref{e:type2b}) is briefly addressed.
It is this particular moment which appears in a natural way in cross sections because 
$E^q$ is accompanied by a factor $\Delta_T$ in the correlator~(\ref{e:qgpd1}), 
while the Sivers function in the correlator~(\ref{e:qtmd1}) is multiplied by $k_T$. 

It is worthwhile to mention that all relations presented in this subsection,
exactly like the relation in~(\ref{e:transsiv3}), also allow one to fix 
(in a model-dependent way) the absolute sign of the Sivers effect of light quarks in 
the nucleon.
(Compare the corresponding discussion in Sec.~\ref{s:four}\,B.)

\subsection{Relations of third type}
\begin{figure*}[t]
 \includegraphics{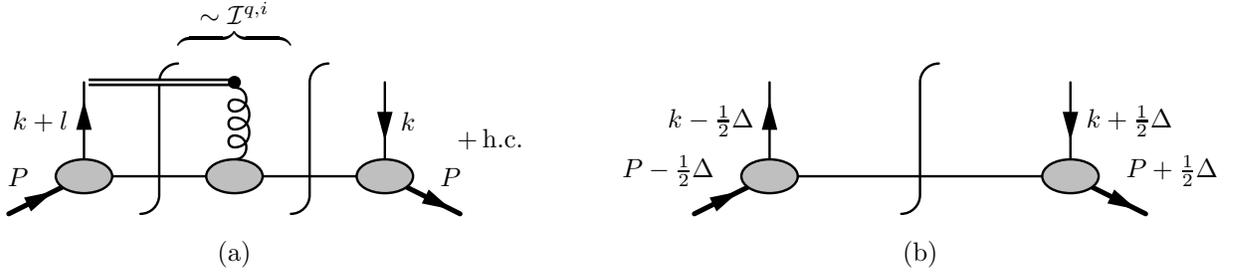}
 \caption{(a) Lowest order diagram for T-odd TMDs in spectator model calculations containing the 
 interaction of the active quark with the target remnant. 
 The eikonal propagator arising from the Wilson line in the operator definition of TMDs
 is indicated by a double line.
 Note that only the imaginary part of the box diagram on the left-hand side (LHS) of the cut is 
 relevant for the calculation of T-odd functions.
 The Hermitian conjugate diagram (h.c.) is not shown.
 (b) Lowest order diagram for GPDs in spectator model calculations.
 The topology of diagram (a) matches with the one of diagram (b) if the quark-spectator
 interaction, described by the lensing function $\mathcal{I}^{q,i}$, is factored out.} 
 \label{f:4}
\end{figure*}
The discussion in Sec.~\ref{s:four}\,B and~\ref{s:four}\,C was exclusively devoted to 
the relations of second type.
Here we want to consider the relations of third type indicated in~(\ref{e:type3}) 
and~(\ref{e:type3n}).
As already pointed out in Sec.~\ref{s:three}\,C one can expect the specific form 
of possible relations of third type to be different from the relations of second 
type.
This expectation is also supported by the fact that all the TMDs 
in~(\ref{e:type2}) and~(\ref{e:type2n}) are T-odd, while in~(\ref{e:type3}) and~(\ref{e:type3n}) 
they are T-even.
To the best of our knowledge an explicit form of a relation of third type does not 
exist in the literature.

Analogous to the result in~(\ref{e:type2mom}) a general relation between the moments of 
TMDs and the moments of GPDs can be established.
To be specific the diquark spectator model results for $h_{1T}^{\bot q}$ in~(\ref{e:sdmtmd8}) 
and for $\tilde{H}_T^q$ in~(\ref{e:sdmgpd6}) obey 
\begin{align} \label{e:type3mom}
 &h_{1T}^{\bot q\,(n)}(x)\notag\\
 &\ =-\frac{g^2(1-x)}{4(2\pi)}\,\frac{n\tilde M^{2n-2}(x)}{2^nM^{2n-2}\,\sin(n\pi)}\notag\\
 &\ =\frac{1}{(2\pi)\,(1-x)^2}\,\frac{n\,\Gamma(4-2n)}{\Gamma^2(2-n)}\,\tilde H_T^{q\,(n)}(x) \,,
\end{align}
which again holds for $0\leq n \leq 1$.
The corresponding quark distributions in the quark target model fulfill exactly
the same relation.
Notice that here no replacement of the type~(\ref{e:couprep}) is needed when making the 
transition from the scalar diquark model to the quark target model because the T-even 
TMDs and the GPDs in Eq.~(\ref{e:type3mom}) receive the first nonzero contribution 
at the same order in perturbation theory.
In addition, we find that those gluon distributions in the quark target model, which 
enter the relation of third type as indicated in~(\ref{e:type3n}), satisfy a relation
with exactly the structure of~(\ref{e:type3mom}).

As expected, the general structure of the relation in~(\ref{e:type2mom}) and 
in~(\ref{e:type3mom}) is different.
Note that due to the Wilson line contribution to the T-odd TMDs, the prefactor on 
the RHS in~(\ref{e:type2mom}) contains couplings which do not appear
in~(\ref{e:type3mom}).
Moreover, the relative power of $(1-x)$ between the moments of the TMDs and of the GPDs 
differs for both types of relations.

Evaluating~(\ref{e:type3mom}) for three specific values of $n$ one finds
\begin{align} \label{e:type3a}
 h_{1T}^{\bot q\,(0)}(x) &=\frac{3}{(1-x)^2}\,\tilde H_T^{q}(x,0,0) \,, \\
 \label{e:type3b}
 h_{1T}^{\bot q\,(1/2)}(x) &=\frac{8}{(2\pi)^2\,(1-x)^2}\,\tilde H_T^{q\,(1/2)}(x) \,,\\
 \label{e:type3c}
 h_{1T}^{\bot q\,(1)}(x)& =\frac{1}{(2\pi)\,(1-x)^2}\,\tilde H_T^{q\,(1)}(x) \,.
\end{align}
Equations~(\ref{e:type3a})--(\ref{e:type3c}) are the counterparts of the relations of second 
type in~(\ref{e:type2a})--(\ref{e:type2c}). 

Keeping in mind the discussion in Sec.~\ref{s:three}\,C 
[see in particular~(\ref{e:type3})] one may wonder if the relation of third type 
in~(\ref{e:type3mom}) can be rewritten such that the second derivative of the impact
parameter distribution $\mathcal{\tilde H}_T^{q}$ shows up.
This is indeed possible for arbitrary values of $n$.
Instead of providing a general formula we limit this discussion to the particular 
case $n=1$ in which the most compact and appealing result follows.
To this end we exploit the model-independent identity
\begin{align} \label{e:ident}
& \int d^2\bT\,\frac{\bT^{\,2}}{2M^2}\, 
     2\bigg( \mathcal{\tilde H}_T^{q}(x,\bT^{\,2})\bigg)'' \notag\\
&\ = -\pi \int_0^\infty d b_T^2 \,\frac{1}{2M^2}\, 
     2\bigg( \mathcal{\tilde H}_T^{q}(x,\bT^{\,2})\bigg)' \notag\\
&\ = \frac{\pi}{M^2} \, \mathcal{\tilde H}_T^{q}(x,0) \notag\\
&\ = \frac{1}{(2\pi)\,(1-x)^2}\,\tilde H_T^{q\,(1)}(x) \,.
\end{align}
In~(\ref{e:ident}) integration by parts is used in order to perform the first step.
Combining now Eqs.~(\ref{e:type3c}) and~(\ref{e:ident}) one immediately obtains
\begin{align} \label{e:type3quark}
& h_{1T}^{\bot q\,(1)}(x) =
  \int d^2\kT\,\frac{\kT^{\,2}}{2M^2}\,h_{1T}^{\bot q}(x,\kT^{\,2}) \notag \\
&\ = \int d^2\bT\,\frac{\bT^{\,2}}{2M^2}\, 
     2\bigg( \mathcal{\tilde H}_T^{q}(x,\bT^{\,2})\bigg)''\,.
\end{align}
Note that this relation has a strong similarity to the relations of first type in
Eqs.~(\ref{e:trivial1})--(\ref{e:trivial3}).
Exactly the same result holds for the relation of third type containing the gluon
distributions [see~(\ref{e:type3})], i.e.,
\begin{align} \label{e:type3gluon}
& h_{1}^{\bot g\,(1)}(x) =
  \int d^2\kT\,\frac{\kT^{\,2}}{2M^2}\,h_{1}^{\bot g}(x,\kT^{\,2}) \notag \\
&\ =\int d^2\bT\,\frac{\bT^{\,2}}{2M^2}\, 
    2\bigg(\mathcal{E}_T^g(x,\bT^{\,2})+2\mathcal{\tilde H}_T^g(x,\bT^{\,2})\bigg)'' \,.
\end{align}

\subsection{Relation of fourth type}
Eventually, the relation of fourth type indicated in~(\ref{e:type4}) and~(\ref{e:type4n}) 
is considered.
In the framework of the quark target model calculation at lowest order such 
a relation is satisfied because both the TMD $h_{1T}^{\bot g}$ and the GPD $\tilde{H}_T^g$ 
vanish [see Eqs.~(\ref{e:qtmtmd16}) and~(\ref{e:qtmgpd12})].
In order to obtain nonzero results for those distributions higher order diagrams have 
to be studied.
At present one can say neither if higher order results obey a relation of fourth type
nor how the specific form of such a relation could look like.
One can only speculate that a possible relation of fourth type may be similar to the 
relation of second type because in both cases a T-odd TMD enters.

\subsection{Higher order diagrams}
\begin{figure*}[t]
 \includegraphics{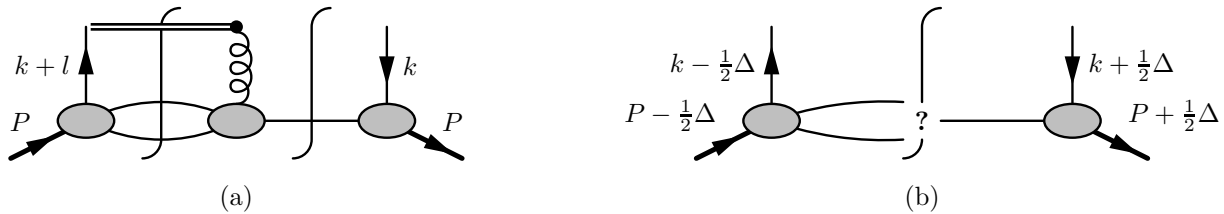}
 \caption{(a) Particular higher order diagram for T-odd TMDs in spectator model calculations containing 
 the interaction of the quark with the target remnants.
 (b) Topology of diagram (a) if the quark-spectator interaction is factored out.
 Diagram (b) cannot represent a Feynman graph for a GPD because there is a mismatch
 between the number of particles on the LHS and the RHS of the cut.} 
 \label{f:5}
\end{figure*}
As already pointed out above so far nontrivial relations between GPDs and TMDs 
are only established if the respective parton distributions are computed in simple
spectator models to lowest nontrivial order in perturbation theory.
It is therefore natural to ask what happens to the relations if higher order 
diagrams in spectator models are taken into account.
Here we would like to briefly address this point by focusing on the relations of
second type.

Consider for instance the relation between the Sivers function and the GPD $E$ as 
given in Eq.~(\ref{e:transsiv3}). 
[The following reasoning equally well applies to the relations of second type
in~(\ref{e:transbm}) and in~(\ref{e:transgluon}).]
We recall that~(\ref{e:transsiv3}) represents a factorization of the Sivers effect
into the distortion of the impact parameter dependent distribution of unpolarized 
quarks in a transversely polarized target, given by the derivative of the GPD 
$\mathcal{E}^q$, times the final state interaction of the active quark, described 
by the lensing function $\mathcal{I}^{q,i}$.

Pictorially this factorization of the Sivers effect is indicated in 
Fig.~\ref{f:4}. 
In order to compute the Sivers function to lowest nontrivial order in a spectator 
model one has to evaluate the cut-diagram in Fig.~\ref{f:4}(a).
If in this diagram the quark-spectator interaction (lensing function) is factored out, 
the topology of the remainder coincides with the diagram in Fig.~\ref{f:4}(b), 
which represents a lowest order Feynman graph for a GPD. 
Hence, it is at least plausible that a factorization as given in Eq.~(\ref{e:transsiv3}) 
can exist.

Suppose now that higher order diagrams are taken into account for the calculation
of the Sivers function, where a particular graph is depicted in Fig.~\ref{f:5}(a).
Factoring out the quark-spectator interaction in this diagram, one ends up with the 
topology shown in Fig.~\ref{f:5}(b).
The diagram in Fig.~\ref{f:5}(b), however, does not represent a Feynman graph for
a GPD, because the number of particles on the LHS and the RHS of the cut
does not match.
Therefore, one can expect that even in spectator models the relations of second type 
no longer hold as soon as higher order diagrams are considered.
Of course, strictly speaking our qualitative discussion here only suggests a breakdown
of these relations and does not provide a rigorous proof.
In any case, in order to arrive at a definite answer on this question a field-theoretic 
definition of the lensing function is needed.
So far no such definition exists.

\section{Summary}
\label{s:five}
Over the last few years several articles appeared (most notably 
Refs.~\cite{Burkardt:2002ks,Burkardt:2003uw,Burkardt:2003je,Diehl:2005jf,Burkardt:2005hp,Brodsky:2006ha,Lu:2006kt}) 
suggesting nontrivial relations between generalized and transverse momentum 
dependent parton distributions.
The present work is dealing with this interesting topic and has mainly a twofold 
purpose:
first, a review of the current knowledge on nontrivial relations between GPDs and
TMDs is given.
Second, the existing results on such relations are considerably extended.

In the following the new results of our work are listed.
\begin{enumerate}
\item The correlator for the chiral-odd leading twist gluon GPDs in impact parameter
 space is written down for the first time [see Eq.~(\ref{e:impact4})].
\item A definition of leading twist gluon TMDs is presented which is very similar to 
 the one proposed in~\cite{Mulders:2000sh} and completely analogous to the definition 
 of quark TMDs in~\cite{Mulders:1995dh,Boer:1997nt,Goeke:2005hb,Bacchetta:2006tn}.
\item In the spirit of Ref.~\cite{Diehl:2005jf} nontrivial relations/analogies between 
 gluon GPDs and TMDs are obtained by comparing the correlators for GPDs in impact 
 parameter representation with the corresponding TMD correlators.
 This procedure allows one to distinguish between different types of relations/analogies, 
 however does not provide an explicit form of a possible relation. 
 The type of the relation is determined, e.g., by the number of derivatives of the GPDs 
 appearing in the correlator in impact parameter representation 
 (see Sec.~\ref{s:three}\,C).
 In our terminology relations of second, third, and fourth type represent nontrivial 
 connections between GPDs and TMDs.
 For instance the relation between the Sivers effect and the GPD $E$ proposed 
 in~\cite{Burkardt:2002ks,Burkardt:2003uw,Burkardt:2003je} is called a relation of 
 second type.
\item It is argued that the momentum space representation of GPDs is also suitable
 in order to find possible relations between GPDs and TMDs.
 (see Sec.~\ref{s:three}\,D).
\item The first calculation of all leading twist GPDs and TMDs in both the scalar 
 diquark model of the nucleon and the quark target model is performed (to lowest 
 nontrivial order in perturbation theory).
\item All our model results, in particular, also those for the gluon distributions,
 obey the relation of second type 
 [see Eqs.~(\ref{e:transsiv3}),~(\ref{e:transbm}), and~(\ref{e:transgluon})], where the 
 specific form of this relation was first given in Ref.~\cite{Burkardt:2003je}.
\item New relations (of second type) for moments of TMDs and GPDs in momentum space 
 (see Sec.~\ref{s:four}\,C) are found, which contain results presented 
 in Refs.~\cite{Burkardt:2003je,Lu:2006kt} as limiting cases.
\item The first explicit form for the relations of third type is given.
 All our model results fulfill this relation [see Sec.~\ref{s:four}\,D, in particular, 
 the compact results in Eqs.~(\ref{e:type3quark}) and~(\ref{e:type3gluon})].
\item The results for the gluon distributions in the quark target model (trivially) 
 satisfy the relation of fourth type in the sense that all involved distributions
 vanish at the order in perturbation theory considered in our work.
\item It is pointed out that the relations of second type are likely to break 
 down in spectator models if higher order perturbative corrections are taken into 
 consideration.
\end{enumerate}
In very brief terms the general status of nontrivial relations between GPDs and TMDs 
can be summarized in the following way:
so far no model-independent nontrivial relations exist and it seems even unlikely 
that they can ever be established.
On the other hand, many relations exist in the framework of simple spectator models.
The phenomenology and the predictive power of the spectator model relation between
the Sivers effect and the GPD $E$ works quite well.
This is the only relation which so far has been tested to some extent.
Additional input from both the experimental and theoretical side is required in
order to further study all other relations between GPDs and TMDs.
Future work will certainly shed more light on this interesting topic.

\section*{ACKNOWLEDGMENTS}
Discussions with M.~Burkardt and M.~Diehl are gratefully acknowledged.
This research is part of the EU Integrated Infrastructure Initiative
Hadronphysics Project under Contract Number RII3-CT-2004-506078.
This work is partially supported by the Verbundforschung ``Hadronen und
Kerne'' of the BMBF.

\onecolumngrid
\appendix
\section{Scalar diquark model}
\label{app:a}
This appendix contains some elements of the scalar diquark model of the 
nucleon (see, e.g., Ref.~\cite{Brodsky:2002cx}) and, in particular, the results of the 
various parton distributions in that approach.
Since in the diquark model not only quarks but also diquarks are considered
as elementary fields we also provide the definition of GPDs and TMDs for scalar 
partons in a spin-$\tfrac{1}{2}$ hadron.

The Lagrangian of the diquark model reads
\begin{align}
 \mathcal{L_\text{SDM}}(x)
 &=\bar\Psi(x)\,(i\gamma^\mu\partial_\mu-M)\,\Psi(x)
  +\bar\psi(x)\,(i\gamma^\mu D_{q,\mu}-m_q)\,\psi(x)
  +\varphi^*(x)\,(\loarrow D_s^{\mu*}\,\roarrow D_{s,\mu}-m_s^2)\,\varphi(x)\notag\\
 &\quad\,-\tfrac{1}{4}F^{\mu\nu}(x)\,F_{\mu\nu}(x)
  +g\,\big[\bar\psi(x)\,\Psi(x)\,\varphi^*(x)+\bar\Psi(x)\,\psi(x)\,\varphi(x)\big] \,,
\end{align}
where $\Psi$ denotes the fermionic target field, $\psi$ the quark field, and $\varphi$ 
the scalar diquark field. 
The essential ingredient of the model is a three-point vertex between the target, quarks, 
and diquarks, with the coupling constant $g$.
This framework allows one to carry out perturbative calculations.
All the results for parton distributions given below contain the coupling $g$ to
the second power, which is the lowest nontrivial order.

In its simplest form the diquark model is of Abelian nature, where both quarks and 
scalar diquarks couple to a photon field via the covariant derivatives
\begin{align}
 D_q^\mu\,\psi(x)=\big[\partial^\mu+ie_q\,A^\mu(x)\big]\,\psi(x) \,, \qquad
 D_s^\mu\,\varphi(x)=\big[\partial^\mu+ie_s\,A^\mu(x)\big]\,\varphi(x) \,,
\end{align}
with the charges $e_q$ and $e_s$, respectively. 
In this model the target has no (electromagnetic) charge reflecting the fact that in QCD 
a hadron is color neutral.
This condition directly implies $e_q=-e_s$\,.
The field strength tensor of the photon is defined in the standard way by
\begin{equation}
 F^{\mu\nu}(x)=\partial^\mu A^\nu(x)-\partial^\nu A^\mu(x) \,.
\end{equation}
Eventually, we mention that the condition $M<m_q+m_s$
has to be fulfilled in order to have a stable target state.

For scalar fields only two GPDs ($H^s$,~$E^s$) and two TMDs ($f_1^s$,~ $f_{1T}^{\bot s}$)
exist.
In analogy with Eqs.~(\ref{e:qgpd}) and~(\ref{e:qgpd1}) one defines the leading twist
GPDs for scalars according to
\begin{align}
 &F^s(x,\Delta;\lambda,\lambda')\notag\\
 &\ =xP^+\int\frac{dz^-}{2\pi}\,e^{i k\cdot z}\,
  \big<p';\lambda'\big|\,
  \varphi^*\big(\!-\!\tfrac{1}{2}z\big)\,
  \mathcal{W}\big(\!-\!\tfrac{1}{2}z;\tfrac{1}{2}z\big)\,
  \varphi\big(\tfrac{1}{2}z\big)\,\big|p;\lambda\big>\,
  \Big|\,\!_{\substack{z^+=0^+\\\zT=\nT}}\notag\\
 &\ =\frac{1}{2P^+}\,\bar u(p',\lambda')\,
  \bigg(\gamma^+\,H^s(x,\xi,t)
  +\frac{i\sigma^{+\mu}\Delta_\mu}{2M}\,E^s(x,\xi,t)\bigg)\,
  u(p,\lambda) \,.
\end{align}
The correlator in impact parameter representation (at $\xi=0$) is given by
\begin{align}
 \mathcal{F}^s(x,\bT;S)=\mathcal{H}^s(x,\bT^{\,2})
  +\frac{\epsilon_T^{ij}b_T^iS_T^j}{M}\,\bigg(\mathcal{E}^s(x,\bT^{\,2})\bigg)' \,,
\end{align}
i.e., it coincides in its form with Eq.~(\ref{e:impact1}).
The definition of TMDs for scalar fields reads
\begin{align}
 &\Phi^s(x,\kT;S)\notag\\
 &\ =xP^+\int\frac{dz^-}{2\pi}\,\frac{d^2\zT}{(2\pi)^2}\,e^{i k\cdot z}\,
  \big<P;S\big|\,
  \varphi^*\big(\!-\!\tfrac{1}{2}z\big)\,
  \mathcal{W}_{+\infty}\big(\!-\!\tfrac{1}{2}z;\tfrac{1}{2}z\big)\,
  \varphi\big(\tfrac{1}{2}z\big)\,\big|P;S\big>\,
  \Big|_{z^+=0^+}\notag\\
 &\ =f_1^s(x,\kT^{\,2})
  -\frac{\epsilon^{ij}_T k_T^i S_T^j}{M}\,f_{1T}^s(x,\kT^{\,2}) \,,
\end{align}
and is analogous to Eqs.~(\ref{e:qtmd}) and~(\ref{e:qtmd1}).
We also note that in the diquark model an obvious change in the definition of the Wilson 
lines in the different parton correlation functions appears: instead of the strong 
coupling $g$ the charges $-e_q$ and $-e_s$ have to be used in the 
correlators of quark and scalar diquark distributions, respectively.
\begin{figure*}
 \includegraphics{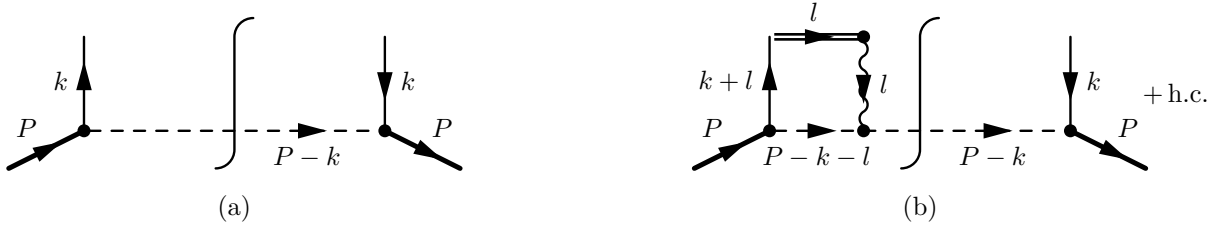}
 \caption{Cut diagrams contributing to quark TMDs and GPDs in the scalar diquark model:
  (a) diagram for T-even TMDs and GPDs; (b) diagram for T-odd TMDs.
  Note that in the case of GPDs the kinematics of Fig.~\ref{f:1} has to be used.}
 \label{f:6}
\end{figure*}
\begin{figure*}
 \includegraphics{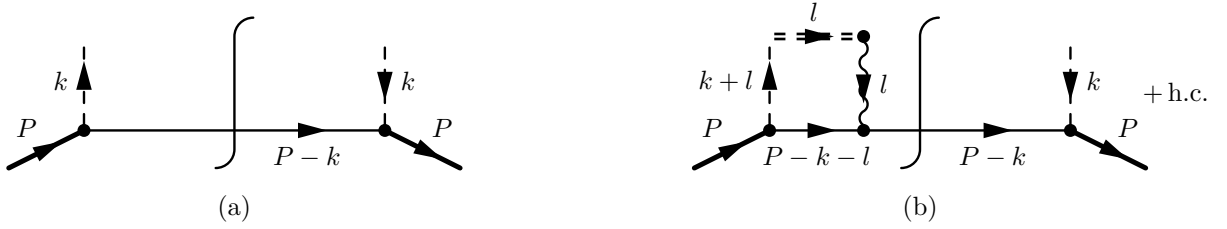}
 \caption{Cut diagrams contributing to diquark TMDs and GPDs in the scalar diquark model:
  (a) diagram for T-even TMDs and GPDs; (b) diagram for T-odd TMDs.
  Note that in the case of GPDs the kinematics of Fig.~\ref{f:1} has to be used.}
\label{f:7}
\end{figure*}

The perturbative calculation of the TMDs and GPDs in the scalar diquark model is 
basically straightforward.
The parton distributions are given by the diagrams in Figs.~\ref{f:6} and~\ref{f:7}.
In the following we just quote the final results without providing any details of 
the calculation.
We begin with the TMDs, where our results for T-even functions are of ${\mathcal O}(g^2)$, 
while the first nontrivial results for T-odd functions exist at 
${\mathcal O}(g^2e_qe_s)$.
This difference appears because necessarily a contribution from the gauge link is 
required in order to generate a nonzero T-odd function.
Note that to this order all TMDs for antiquarks vanish identically.
For the eight TMDs of quarks and the two TMDs of scalar diquarks one finds
\begin{align} \label{e:sdmtmd1}
 f_1^q(x,\kT^{\,2})
 &=\frac{g^2(1-x)}{2(2\pi)^3}\,
  \frac{\kT^{\,2}+(m_q+xM)^2}
  {\big[\kT^{\,2}+\tilde M^2(x)\big]\,\!^2} \,,
 \displaybreak[0]\\ \label{e:sdmtmd2}
 f_{1T}^{\bot q}(x,\kT^{\,2})
 &=\frac{g^2e_qe_s(1-x)}{4(2\pi)^4}\,
  \frac{(m_q+xM)M}
  {\kT^{\,2}\,\big[\kT^{\,2}+\tilde M^2(x)\big]} \,
  \ln\bigg(\frac{\kT^{\,2}+\tilde M^2(x)}{\tilde M^2(x)}\bigg) \,,
 \displaybreak[0]\\
 g_{1L}^q(x,\kT^{\,2})
 &=-\frac{g^2(1-x)}{2(2\pi)^3}\,
  \frac{\kT^{\,2}-(m_q+xM)^2}
  {\big[\kT^{\,2}+\tilde M^2(x)\big]\,\!^2} \,,
 \displaybreak[0]\\
 g_{1T}^q(x,\kT^{\,2})
 &=\frac{g^2(1-x)}{(2\pi)^3}\,
  \frac{(m_q+xM)M}
  {\big[\kT^{\,2}+\tilde M^2(x)\big]\,\!^2} \,,
 \displaybreak[0]\\ \label{e:sdmtmd5}
 h_1^{\bot q}(x,\kT^{\,2})
 &=\frac{g^2e_qe_s(1-x)}{4(2\pi)^4}\,
  \frac{(m_q+xM)M}
  {\kT^{\,2}\,\big[\kT^{\,2}+\tilde M^2(x)\big]} \,
  \ln\bigg(\frac{\kT^{\,2}+\tilde M^2(x)}{\tilde M^2(x)}\bigg) \,,
 \displaybreak[0]\\
 h_{1L}^{\bot q}(x,\kT^{\,2})
 &=-\frac{g^2(1-x)}{(2\pi)^3}\,
  \frac{(m_q+xM)M}
  {\big[\kT^{\,2}+\tilde M^2(x)\big]\,\!^2} \,,
 \displaybreak[0]\\
 h_{1T}^q(x,\kT^{\,2})
 &=\frac{g^2(1-x)}{2(2\pi)^3}\,
  \frac{\kT^{\,2}+(m_q+xM)^2}
  {\big[\kT^{\,2}+\tilde M^2(x)\big]\,\!^2} \,,
 \displaybreak[0]\\ \label{e:sdmtmd8}
 h_{1T}^{\bot q}(x,\kT^{\,2})
 &=-\frac{g^2(1-x)}{(2\pi)^3}\,
  \frac{M^2}
  {\big[\kT^{\,2}+\tilde M^2(x)\big]\,\!^2} \,,
 \displaybreak[0]\\
 f_1^s(x,\kT^{\,2})
 &=\frac{g^2x}{2(2\pi)^3}\,
  \frac{\kT^{\,2}+\big(m_q+(1-x)M\big)\,\!^2}
  {\big[\kT^{\,2}+\tilde M^2(1-x)\big]\,\!^2} \,,
 \displaybreak[0]\\ \label{e:sdmtmd10}
 f_{1T}^{\bot s}(x,\kT^{\,2})
 &=-\frac{g^2e_qe_sx}{4(2\pi)^4}\,
  \frac{\big(m_q+(1-x)M\big)M}
  {\kT^{\,2}\,\big[\kT^{\,2}+\tilde M^2(1-x)\big]} \,
  \ln\bigg(\frac{\kT^{\,2}+\tilde M^2(1-x)}{\tilde M^2(1-x)}\bigg) \,.
\end{align}
In the above formulas we used the abbreviation
\begin{equation}
 \tilde M^2(x)=(1-x)m_q^2+xm_s^2-x(1-x)M^2 
\end{equation}
for some specific combination of mass terms.
All results in~(\ref{e:sdmtmd1})--(\ref{e:sdmtmd10}) were already given in the 
literature.
To be specific, the full list of T-even quark TMDs was computed in 
Ref.~\cite{Jakob:1997wg}.
The quark Sivers function was considered in~\cite{Collins:2002kn,Ji:2002aa} 
(see also~\cite{Bacchetta:2003rz}), and the Boer-Mulders function 
in~\cite{Goldstein:2002vv,Boer:2002ju}.
The result for the Sivers function of the scalar diquark can be found in 
Ref.~\cite{Goeke:2006ef}.

Now we proceed to the model results for the GPDs, where in all cases nonzero results
are obtained at ${\mathcal O}(g^2)$.
Again, to this order all GPDs for antiquarks vanish identically.
We limit ourselves to the case $\xi=0$ which is sufficient for the purpose of our work.
The results read 
\begin{align} \label{e:sdmgpd1}
 H^q(x,0,-\DT^2)
 &=\frac{g^2(1-x)}{2(2\pi)^3}\,\int d^2\kT\,
  \frac{\kT^{\,2}-\tfrac{1}{4}(1-x)^2\DT^2+(m_q+xM)^2}
  {D_{\textrm{SDM}}^q(x,\DT;\kT)} \,,
 \displaybreak[0]\\ \label{e:sdmgpd2}
 E^q(x,0,-\DT^2)
 &=\frac{g^2(1-x)^2}{(2\pi)^3}\,\int d^2\kT\,
  \frac{(m_q+xM)M}
  {D_{\textrm{SDM}}^q(x,\DT;\kT)} \,,
 \displaybreak[0]\\
 \tilde H^q(x,0,-\DT^2)
 &=-\frac{g^2(1-x)}{2(2\pi)^3}\,\int d^2\kT\,
  \frac{\kT^{\,2}-\tfrac{1}{4}(1-x)^2\DT^2-(m_q+xM)^2}
  {D_{\textrm{SDM}}^q(x,\DT;\kT)} \,,
 \displaybreak[0]\\
 H_T^q(x,0,-\DT^2)
 &=-\frac{g^2(1-x)}{2(2\pi)^3}\,\int d^2\kT\,
  \frac{\tfrac{\DT^2}{\Delta_T^1\Delta_T^2}\big(k_T^1k_T^2
  -\tfrac{1}{4}(1-x)^2\Delta_T^1\Delta_T^2\big)-(m_q+xM)^2}
  {D_{\textrm{SDM}}^q(x,\DT;\kT)} \,,
 \displaybreak[0]\\ \label{e:sdmgpd5}
 E_T^q(x,0,-\DT^2)
 &=-\frac{g^2(1-x)}{(2\pi)^3}\,\int d^2\kT\,
  \frac{\tfrac{4M^2}{\Delta_T^1\Delta_T^2}\big(k_T^1k_T^2-\tfrac{1}{4}(1-x)^2\Delta_T^1\Delta_T^2\big)-(1-x)(m_q+xM)M}
  {D_{\textrm{SDM}}^q(x,\DT;\kT)} \,,
 \displaybreak[0]\\ \label{e:sdmgpd6}
 \tilde H_T^q(x,0,-\DT^2)
 &=\frac{2g^2(1-x)}{(2\pi)^3}\,\int d^2\kT\,
  \frac{\tfrac{M^2}{\Delta_T^1\Delta_T^2}\big(k_T^1k_T^2-\tfrac{1}{4}(1-x)^2\Delta_T^1\Delta_T^2\big)}
  {D_{\textrm{SDM}}^q(x,\DT;\kT)} \,,
 \displaybreak[0]\\
 H^s(x,0,-\DT^2)
 &=\frac{g^2x}{2(2\pi)^3}\,\int d^2\kT\,
  \frac{\kT^{\,2}-\tfrac{1}{4}(1-x)^2\DT^2+\big(m_q+(1-x)M\big)\,\!^2}
  {D_{\textrm{SDM}}^s(x,\DT;\kT)} \,,
 \displaybreak[0]\\ \label{e:sdmgpd8}
 E^s(x,0,-\DT^2)
 &=-\frac{g^2x(1-x)}{(2\pi)^3}\,\int d^2\kT\,
  \frac{\big(m_q+(1-x)M\big)M}
  {D_{\textrm{SDM}}^s(x,\DT;\kT)} \,,
\end{align}
with the denominators
\begin{align}
 {D_{\textrm{SDM}}^q(x,\DT;\kT)}&=
  \big[\big(\kT-\tfrac{1}{2}(1-x)\DT\big)\,\!^2+\tilde M^2(x)\big]\,
  \big[\big(\kT+\tfrac{1}{2}(1-x)\DT\big)\,\!^2+\tilde M^2(x)\big] \,, \\
 {D_{\textrm{SDM}}^s(x,\DT;\kT)}&=
  \big[\big(\kT-\tfrac{1}{2}(1-x)\DT\big)\,\!^2+\tilde M^2(1-x)\big]\,
  \big[\big(\kT+\tfrac{1}{2}(1-x)\DT\big)\,\!^2+\tilde M^2(1-x)\big] \,.
\end{align}
We refrain from performing the integration upon the transverse quark momentum in the 
results given in~(\ref{e:sdmgpd1})--(\ref{e:sdmgpd8}) because this step is actually not 
needed for studying the relations between GPDs and TMDs. 
Note also that in the case of $H$ and $\tilde{H}$ the $k_T$ integration leads to a 
(logarithmic) ultraviolet divergence, which is not a peculiarity of the diquark 
model but rather a well-known feature of light-cone correlation functions.
For the other GPDs this integral is finite, which for $E$ is directly obvious and
in the remaining cases can be shown by explicit calculation.

\section{Quark target model}
\label{app:b}
The second model we are using is a quark target, treated in perturbative QCD.
Mainly for two reasons this approach is of interest in the context of our investigation.
First, the non-Abelian three-gluon vertex now enters the computation of various parton
distributions. 
Therefore, one has an additional check of relations between GPDs and TMDs, beyond what 
can be done in the Abelian scalar diquark model.
Second, the quark target model allows one to investigate relations between gluon 
distributions.

The Lagrangian of the quark target model is given by
\begin{align} \label{e:lagqtm}
 \mathcal{L}_\text{QTM}(x)
 =\bar\psi(x)\,(i\gamma^\mu D_\mu-m)\,\psi(x)
 -\tfrac{1}{4}F^{\mu\nu}_a(x)\,F_{\mu\nu,a}(x) \,,
\end{align}
i.e., it coincides the QCD Lagrangian for the specific case of a single quark 
flavor.
The covariant derivative in~(\ref{e:lagqtm}) reads
\begin{equation}
 D^\mu\,\psi(x)=\big[\partial^\mu-ig\,t_a\,A^\mu_a(x)\big]\,\psi(x) \,,
\end{equation}
while the field strength tensor is defined in Eq.~(\ref{e:fqcd}).

\begin{figure*}
 \includegraphics{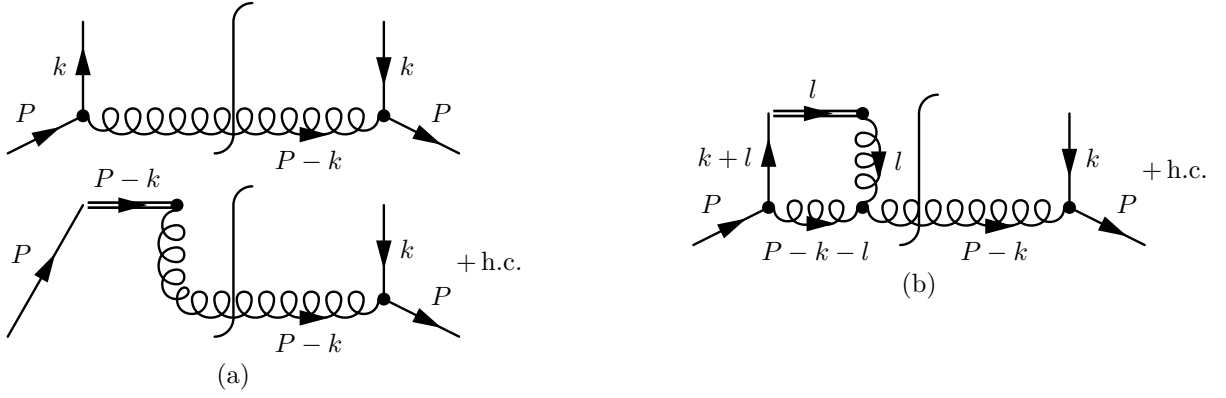}
 \caption{Cut diagrams contributing to quark TMDs and GPDs in the quark target model:
 (a) diagrams for T-even TMDs and GPDs;
 (b) diagram for T-odd TMDs.
 Note that in the case of GPDs the kinematics of Fig.~\ref{f:1} has to be used.}
\label{f:8}
\end{figure*}

\begin{figure*}
 \includegraphics{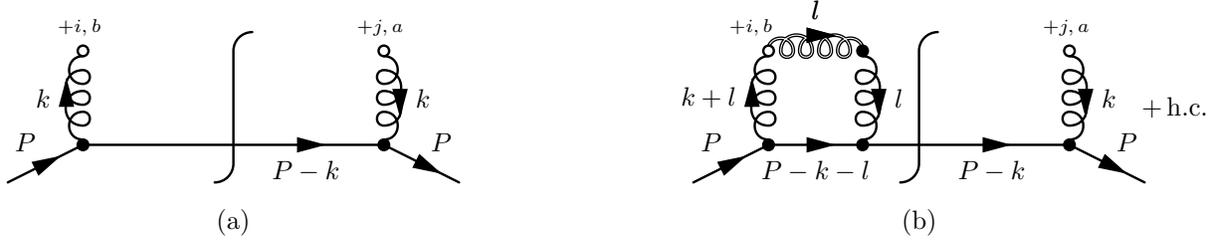}
 \caption{Cut diagrams contributing to gluon TMDs and GPDs in the quark target model:
 (a) diagram for T-even TMDs and GPDs; (b) diagram for T-odd TMDs.
 Note that in the case of GPDs the kinematics of Fig.~\ref{f:1} has to be used,
 and that diagram (b) contains a Wilson line for gluons.
 The open circle on the upper end of the gluon lines indicates a special Feynman rule
 for the field strength tensor in the definition of parton
 distributions for gluons (see Refs.~\cite{Collins:1981uw,Goeke:2006ef}).}
\label{f:9}
\end{figure*}

We now list the results for TMDs and GPDs in the quark target model.
Like in the case of the diquark model of the nucleon here we also avoid giving any 
details of the calculation.
The relevant diagrams for the quark distributions are depicted in Fig.~\ref{f:8},
and those for the gluon distributions in Fig.~\ref{f:9}.
The T-odd TMDs receive the first nonzero contribution at ${\mathcal O}(g^4)$, while
the results for all other distributions (T-even TMDs as well as GPDs) are of
${\mathcal O}(g^2)$.

For the eight TMDs of quarks and the eight TMDs of gluons one finds (in the kinematic
region $0<x<1$),
\begin{align} \label{e:qtmtmd1}
 f_1^q(x,\kT^{\,2})
 &=\frac{4g^2}{3(2\pi)^3(1-x)}\,
  \frac{\big(1+x^2\big)\kT^{\,2}+(1-x)^4\,m^2}
  {\big[\kT^{\,2}+(1-x)^2\,m^2\big]\,\!^2} \,,
 \displaybreak[0]\\
 f_{1T}^{\bot q}(x,\kT^{\,2})
 &=-\frac{g^4x(1-x)}{(2\pi)^4}\,
  \frac{m^2}
  {\kT^{\,2}\,\big[\kT^{\,2}+(1-x)^2\,m^2\big]} \,
  \ln\bigg(\frac{\kT^{\,2}+(1-x)^2\,m^2}{(1-x)^2\,m^2}\bigg) \,,
 \displaybreak[0]\\
 g_{1L}^q(x,\kT^{\,2})
 &=\frac{4g^2}{3(2\pi)^3(1-x)}\,
  \frac{\big(1+x^2\big)\kT^{\,2}-(1-x)^4\,m^2}
  {\big[\kT^{\,2}+(1-x)^2\,m^2\big]\,\!^2} \,,
 \displaybreak[0]\\
 g_{1T}^q(x,\kT^{\,2})
 &=-\frac{8g^2x(1-x)}{3(2\pi)^3}\,
  \frac{m^2}
  {\big[\kT^{\,2}+(1-x)^2\,m^2\big]\,\!^2} \,,
 \displaybreak[0]\\
 h_1^{\bot q}(x,\kT^{\,2})
 &=-\frac{g^4(1-x)}{(2\pi)^4}\,
  \frac{m^2}
  {\kT^{\,2}\,\big[\kT^{\,2}+(1-x)^2\,m^2\big]} \,
  \ln\bigg(\frac{\kT^{\,2}+(1-x)^2\,m^2}{(1-x)^2\,m^2}\bigg) \,,
 \displaybreak[0]\\
 h_{1L}^{\bot q}(x,\kT^{\,2})
 &=\frac{8g^2(1-x)}{3(2\pi)^3}\,
  \frac{m^2}
  {\big[\kT^{\,2}+(1-x)^2\,m^2\big]\,\!^2} \,,
 \displaybreak[0]\\
 h_{1T}^q(x,\kT^{\,2})
 &=\frac{8g^2x}{3(2\pi)^3(1-x)}\,
  \frac{\kT^{\,2}}
  {\big[\kT^{\,2}+(1-x)^2\,m^2\big]\,\!^2} \,,
 \displaybreak[0]\\
 h_{1T}^{\bot q}(x,\kT^{\,2})
 &=0 \,,
 \displaybreak[0]\\
 f_1^g(x,\kT^{\,2})
 &=\frac{4g^2}{3(2\pi)^3x}\,
  \frac{\big(1+(1-x)^2\big)\kT^{\,2}+x^4m^2}
  {\big[\kT^{\,2}+x^2m^2\big]\,\!^2} \,,
 \displaybreak[0]\\
 f_{1T}^{\bot g}(x,\kT^{\,2})
 &=\frac{g^4x(1-x)}{(2\pi)^4}\,
  \frac{m^2}
  {\kT^{\,2}\,\big[\kT^{\,2}+x^2m^2\big]} \,
  \ln\bigg(\frac{\kT^{\,2}+x^2m^2}{x^2m^2}\bigg) \,,
 \displaybreak[0]\\
 g_{1L}^g(x,\kT^{\,2})
 &=\frac{4g^2}{3(2\pi)^3}\,
  \frac{\big(1+(1-x)\big)\kT^{\,2}+x^3m^2}
  {\big[\kT^{\,2}+x^2m^2\big]\,\!^2} \,,
 \displaybreak[0]\\
 g_{1T}^g(x,\kT^{\,2})
 &=-\frac{8g^2x(1-x)}{3(2\pi)^3}\,
  \frac{m^2}
  {\big[\kT^{\,2}+x^2m^2\big]\,\!^2} \,,
 \displaybreak[0]\\
 h_1^{\bot g}(x,\kT^{\,2})
 &=\frac{16g^2(1-x)}{3(2\pi)^3x}\,
  \frac{m^2}
  {\big[\kT^{\,2}+x^2m^2\big]\,\!^2} \,,
 \displaybreak[0]\\
 h_{1L}^{\bot g}(x,\kT^{\,2})
 &=0 \,,
 \displaybreak[0]\\ \label{e:qtmtmd15}
 h_{1T}^g(x,\kT^{\,2})
 &=\frac{2g^4x}{(2\pi)^4}\,
  \frac{m^2}
  {\kT^{\,2}\,\big[\kT^{\,2}+x^2m^2\big]} \,
  \ln\bigg(\frac{\kT^{\,2}+x^2m^2}{x^2m^2}\bigg) \,,
 \displaybreak[0]\\ \label{e:qtmtmd16}
 h_{1T}^{\bot g}(x,\kT^{\,2})
 &=0 \,.
\end{align}
At the kinematical point $x=1$ there exist additional contributions for some of 
the parton distributions above, which arise if the on-shell intermediate state in the 
cut diagram is just the vacuum.
Those contributions are neglected here for simplicity.
Note that three TMDs in (\ref{e:qtmtmd1})--(\ref{e:qtmtmd16}) vanish.
It is likely that a higher order calculation provides a nonzero result for these objects.
Some of the results in~(\ref{e:qtmtmd1})--(\ref{e:qtmtmd16}) were already given in the 
literature.
Treatments of the quark TMDs $f_1^q,\;g_{1T}^q,\;h_{1L}^{\bot q},\;h_{1T}^q$ can for 
instance be found 
in~\cite{Mukherjee:2001zx,Kundu:2001pk,Collins:2003fm,Schlegel:2004rg,Hautmann:2007uw}.
The Sivers function for quarks and gluons was computed in Ref.~\cite{Goeke:2006ef}.

Eventually, we consider the GPDs in the quark target model.
The results (for $\xi = 0$ and $0<x<1$) read
\begin{align} \label{e:qtmgpd1}
 H^q(x,0,-\DT^2)
 &=\frac{4g^2}{3(2\pi)^3(1-x)}\,\int d^2\kT\,
  \frac{\big(1+x^2\big)\big(\kT^{\,2}-\tfrac{1}{4}(1-x)^2\DT^2\big)+(1-x)^4\,m^2}
  {D_{\textrm{QTM}}^q(x,\DT;\kT)} \,,
 \displaybreak[0]\\
 E^q(x,0,-\DT^2)
 &=\frac{8g^2x(1-x)^2}{3(2\pi)^3}\,\int d^2\kT\,
  \frac{m^2}
  {D_{\textrm{QTM}}^q(x,\DT;\kT)} \,,
 \displaybreak[0]\\
 \tilde H^q(x,0,-\DT^2)
 &=\frac{4g^2}{3(2\pi)^3(1-x)}\,\int d^2\kT\,
  \frac{\big(1+x^2\big)\big(\kT^{\,2}-\tfrac{1}{4}(1-x)^2\DT^2\big)-(1-x)^4\,m^2}
  {D_{\textrm{QTM}}^q(x,\DT;\kT)} \,,
 \displaybreak[0]\\
 H_T^q(x,0,-\DT^2)
 &=\frac{8g^2x}{3(2\pi)^3(1-x)}\,\int d^2\kT\,
  \frac{\kT^{\,2}-\tfrac{1}{4}(1-x)^2\DT^2}
  {D_{\textrm{QTM}}^q(x,\DT;\kT)} \,,
 \displaybreak[0]\\
 E_T^q(x,0,-\DT^2)
 &=\frac{8g^2(1-x)^2}{3(2\pi)^3}\,\int d^2\kT\,
  \frac{m^2}
  {D_{\textrm{QTM}}^q(x,\DT;\kT)} \,,
 \displaybreak[0]\\
 \tilde H_T^q(x,0,-\DT^2)
 &=0 \,,
 \displaybreak[0]\\
 H^g(x,0,-\DT^2)
 &=\frac{4g^2}{3(2\pi)^3x}\,\int d^2\kT\,
  \frac{\big(1+(1-x)^2\big)\big(\kT^{\,2}-\tfrac{1}{4}(1-x)^2\DT^2\big)+x^4m^2}
  {D_{\textrm{QTM}}^g(x,\DT;\kT)} \,,
 \displaybreak[0]\\
 E^g(x,0,-\DT^2)
 &=-\frac{8g^2x(1-x)^2}{3(2\pi)^3}\,\int d^2\kT\,
  \frac{m^2}
  {D_{\textrm{QTM}}^g(x,\DT;\kT)} \,,
 \displaybreak[0]\\
 \tilde H^g(x,0,-\DT^2)
 &=\frac{4g^2}{3(2\pi)^3}\,\int d^2\kT\,
  \frac{\big(1+(1-x)\big)\big(\kT^{\,2}-\tfrac{1}{4}(1-x)^2\DT^2\big)+x^3m^2}
  {D_{\textrm{QTM}}^g(x,\DT;\kT)} \,,
 \displaybreak[0]\\ \label{e:qtmgpd10}
 H_T^g(x,0,-\DT^2)
 &=-\frac{8g^2x(1-x)}{3(2\pi)^3}\,\int d^2\kT\,
  \frac{m^2}
  {D_{\textrm{QTM}}^g(x,\DT;\kT)} \,,
 \displaybreak[0]\\
 E_T^g(x,0,-\DT^2)
 &=-\frac{32g^2(1-x)}{3(2\pi)^3x}\,\int d^2\kT\,
  \frac{\tfrac{m^2}{\Delta_T^1\Delta_T^2}\,\big(k_T^1k_T^2
   -\tfrac{1}{4}(1-x)^2\Delta_T^1\Delta_T^2\big)}
  {D_{\textrm{QTM}}^g(x,\DT;\kT)} \,,
 \displaybreak[0]\\ \label{e:qtmgpd12}
 \tilde H_T^g(x,0,-\DT^2)
 &=0 \,,
\end{align}
with the denominators
\begin{align}
 {D_{\textrm{QTM}}^q(x,\DT;\kT)}&=
  \big[\big(\kT-\tfrac{1}{2}(1-x)\DT\big)\,\!^2+(1-x)^2\,m^2\big]\,
  \big[\big(\kT+\tfrac{1}{2}(1-x)\DT\big)\,\!^2+(1-x)^2\,m^2\big] \,, \\
 {D_{\textrm{QTM}}^g(x,\DT;\kT)}&=
  \big[\big(\kT-\tfrac{1}{2}(1-x)\DT\big)\,\!^2+x^2m^2\big]\,
  \big[\big(\kT+\tfrac{1}{2}(1-x)\DT\big)\,\!^2+x^2m^2\big] \,.
\end{align}
Calculations of the chiral-even GPDs for both quarks and gluons can be found in
Refs.~\cite{Mukherjee:2002pq,Mukherjee:2002xi,Chakrabarti:2004ci}.
To our knowledge the results for the chiral-odd GPDs 
in~(\ref{e:qtmgpd1})--(\ref{e:qtmgpd12}) are given here for the first time.
\twocolumngrid

\bibliography{gpd_tmd}
\end{document}